\documentclass[11pt, a4paper, preprint, nofootinbib, prd]{revtex4-2}
\usepackage{geometry}
\usepackage{mathtools}
\usepackage{bm,physics}
\usepackage{enumerate}
\usepackage{extarrows}
\usepackage{epsf,amssymb,bbm,amsbsy,amsfonts,amssymb,amsmath,booktabs,mathdots,fancyhdr}
\numberwithin{equation}{section}
\usepackage{hyperref}
\usepackage{graphicx}
\usepackage[normalem]{ulem}
\usepackage{tikz}
\usepackage{pgfplots}
\pgfplotsset{compat=1.13}
\usepackage{academicons}
\usepackage{xcolor}
\usepackage{fontawesome5}

\definecolor{orcidlogocol}{rgb}{0.65, 0.807, 0.223}

\newcommand{\orcid}[1]{$\,$\href{https://orcid.org/#1}{\textcolor{orcidlogocol}{\faOrcid}}}

\usepackage[T1]{fontenc} 

\definecolor{darkgreen}{RGB}{0,102,102}
\definecolor{purple}{RGB}{102,0,102}
\definecolor{darkblue}{RGB}{0,0,102}
\hypersetup{
    colorlinks=true,       
    linkcolor=purple,         
    citecolor=purple,        
    filecolor=blue,     
    urlcolor=darkgreen        
}
\begin{document}

\title{Oscillon formation from preheating in asymmetric inflationary potentials}

\author{Rafid Mahbub \orcid{0000-0003-2665-2798}}
\affiliation{Department of Physics, Gustavus Adolphus College, Saint Peter, MN 56082, USA}
\email{mahbub@gustavus.edu}
\author{Swagat S. Mishra \orcid{0000-0003-4057-145X}}
\affiliation{Center for Astronomy and Particle Theory, School of Physics and Astronomy, University of Nottingham, Nottingham, NG7 2RD, UK}
\email{swagat.mishra@nottingham.ac.uk}

\begin{abstract}
    We investigate the possibility of oscillon formation during the preheating phase of asymmetric inflationary potentials. We analytically establish the existence of oscillon-like solutions for the Klein-Gordon equation for a polynomial potential of the form $V(\phi)=\frac{1}{2}\phi^2+A\phi^3+B\phi^4$ using the small amplitude analysis, which naturally arises as a Taylor expansion of the $\alpha$-attractor E-model for $\phi\ll M_\text{pl}$ and $\alpha\sim\mathcal{O}(1)$. We perform a detailed numerical analysis to study the formation of nonlinear structures in the $\alpha$-attractor E-model using the publicly available lattice simulation code \textsf{$\mathcal{C}\text{osmo}\mathcal{L}\text{attice}$} for parameters in the range $10^{-5}\lesssim\alpha\lesssim 5\times 10^{-4}$. We find the  backreaction of the field fluctuations onto the evolution of the homogeneous inflaton condensate to be significant for $\alpha\lesssim 2\times 10^{-4}$ for which we observe the formation of highly nonlinear structures with average equation of state $w\simeq 0$. These nonlinear structures maybe interpreted as \textit{oscillons}, providing evidence that they can form during the inflaton oscillations around  an asymmetric  potential and are found to be present for the entirety of the runtime of our simulations, comprising $\gtrsim 40\%$ of the total energy density.
\end{abstract}
\maketitle
\newpage
\tableofcontents

\section{Introduction}

Cosmic inflation is at present the leading paradigm that explains the origin of structure in the Universe through vacuum quantum fluctuations during such a period of accelerated expansion of space  \cite{Starobinsky:1979ty,Starobinsky:1980te,Guth:1980zm,Mukhanov:1981xt,Guth:1982ec,Linde:1981mu}. Through the temperature anisotropies of the Cosmic Microwave Background (CMB), we can place constraints on inflationary observables $n_s$ and $r$ that help us rule in favor and against various inflationary models \cite{Planck:2018vyg,Planck:2018jri,Paoletti:2022anb}. Inflation ends shortly after the inflaton leaves the slow-roll regime, after which it decays  into the particles of the Standard Model via reheating. Originally, reheating was studied as a perturbative process where the inflaton particles decay into other particles independently, thermalizing the Universe and triggering the onset of the Hot Big Bang \cite{Kofman:1994rk,Kofman:1996mv,Bassett:2005xm,Allahverdi:2010xz}. However, this perturbative picture ignores collective phenomena that can arise from large and coherent oscillations of the inflaton about the minimum of the potential. For example, Bose condensation effects greatly enhance the rate of the inflaton quanta decay that cannot be captured by perturbative calculations. Of particular importance is the phenomenon of \textit{parametric resonance} (along with \textit{tachyonic resonance}) through which perturbations of the inflaton field and other species can become amplified, leading to copious particle production \cite{Figueroa:2016wxr} --  with the field fluctuations growing as $\delta\phi_{\bm{k}}\propto e^{\mu_{\bm{k}}t}$ and $n_{\bm{k}}\propto e^{2\mu_{\bm{k}}t}$ with a characteristic exponent $\mu_{\bm{k}}$ to be defined later. Description of these phenomena requires a non-perturbative treatment. Such a  non-thermal period  of particle production precedes the usual thermal reheating phase and has been aptly named \textit{preheating} \cite{Kofman:1994rk,Amin:2014eta,Lozanov:2019jxc}. The period of preheating gives rise to very interesting nonlinear phenomena that are of great interest to cosmologists.\\
\indent Generic non-linear scalar field theories can exhibit localized configurations in space which oscillate with time. Such configurations can be obtained in scalar field theories with potentials that open up  away from (are shallower than) a quadratic minimum $\frac{m^2\phi^2}{2}$ such that the inflaton feels the effects of attractive self-interaction as it oscillates about the minimum. Such localized and dense configurations are called \textit{oscillons} \cite{Bogolyubsky:1976nx,Gleiser:1993pt,Kusenko:1997zq,Kasuya:2002zs,Farhi:2007wj}. These soliton-like phenomena were first observed in granular and dissipative media \cite{PhysRevE.53.2972,1996Natur.382..793U}, which subsequently found their place in the study of preheating in inflationary dynamics. Initial studies on oscillons in cosmology were performed in the context of cosmological phase transitions \cite{Copeland:1995fq,Riotto:1995yy,Graham:2006vy}. In more recent times, oscillons have been linked to seeding primordial black holes \cite{Cotner:2018vug,Cotner:2019ykd,Widdicombe:2019woy,Nazari:2020fmk} and primordial gravitational waves \cite{Easther:2006gt,Dufaux:2007pt,Antusch:2016con,Liu:2017hua,Amin:2018xfe,Helfer:2018vtq,Hiramatsu:2020obh} among other interesting phenomena. More recently still, oscillons have found their way into the realm of dark matter, in particular ultra-light dark matter (ULDM). Since oscillons do not have a conserved charge associated with them, they are not technically stable (they decay through the emission of radiation \cite{Fodor:2008du,Hertzberg:2010yz,Salmi:2012ta,Zhang:2020bec}). Nevertheless,  since these quasi-stable objects can be extremely long-lived and, combined with the fact that they can be produced in large-enough abundances relevant to cosmology, they can be used to model the formation of non-linear structure in  ULDM  \cite{Hu:2000ke,Olle:2019kbo,Kawasaki:2019czd,Ferreira:2020fam}.\\
\indent Since oscillon formation requires potentials that are shallower than a quadratic potential around the minimum, there are various inflationary models that can be used as a prototype for their formation. Oscillon formation has been studied in $\phi^6$ theories, both analytically and numerically in Refs. \cite{Amin:2010dc,Amin:2010jq,Amin:2010xe}, where it was shown that their field configuration acquires a flat-top feature around the central regions of the core. Moreover, it has also been studied in axion monodromy potentials in Refs. \cite{Amin:2011hj,Zhou:2013tsa,Lozanov:2014zfa,Sang:2019ndv}, hilltop potentials in Refs. \cite{Antusch:2017vga,Antusch:2019qrr} and in the $\alpha$-attractor T-model in Refs. \cite{Lozanov:2017hjm,Lozanov:2019ylm}. In Ref. \cite{Lozanov:2017hjm}, the authors used the T-model to quantify the differences between $n=1$ and $n>1$ (where the minima are quadratic and non-quadratic respectively at leading order), leading to quite different field configurations as end products. They showed that oscillons can form when $n=1$ such that the time-averaged equation of state parameter $\langle w \rangle \rightarrow 0$. In cases where $n>1$, initially dense transients form which fragment into radiation after a few $e$-folds with $\langle w \rangle \rightarrow \frac{1}{3}$. However, the question of whether oscillons can form in asymmetric potentials has not been addressed adequately in the existing literature. Unlike their symmetric counterparts, asymmetric potentials usually do not open up away from the quadratic minimum on both sides. Hence, the question of whether oscillons do form in such potentials becomes important since the oscillating inflaton might only be experiencing the attractive self-interaction on one side of the minimum.\footnote{Even in the absence of self-interactions, nonlinear structures can form due to the amplification of  metric perturbations that are coupled to the  inflaton fluctuations. For example, the scalar perturbations of  the simplest $\phi^2$ theory (without any self-interaction)  undergo parametric resonant amplification on time scales that are much longer than the oscillon formation time scale in the interacting theory \cite{Johnson:2008se,Jedamzik:2010dq,Mishra:2017ehw,Martin:2020fgl}.} These differences between symmetric and asymmetric  potentials are illustrated in Fig. (\ref{fig:even_odd}).  Moreover, since there are only a few well-known asymmetric  inflationary potentials (\textit{e.g.} the Starobinsky model), performing a comprehensive study into the matter becomes difficult due to the limited number of asymmetric potentials available. However, see Refs.~\cite{Antusch:2016con,Amin:2018xfe} for previous studies on  oscillon formation in asymmetric potentials in the context of production of gravitational waves.
\begin{figure}[t]
    \centering
    \begin{tikzpicture}[scale=0.75, transform shape]
       \begin{axis}[
                    axis lines = center,
                    axis line style = thick,
                    xlabel = \(\frac{\phi}{M}\),
                    ylabel = {\( \frac{V}{m^2 M^2} \)},
                    label style={font=\Large},
                    xtick=\empty,
                    ytick=\empty
                    ]
            \addplot [
                        domain=-1.5:1.5,
                        range=0:2.5,
                        samples=100, 
                        color=black,
                        very thick
                    ]{0.5*x^2};
            \addplot [
                        domain=-1.5:1.5,
                        range=0:2.5,
                        samples=100, 
                        color=gray,
                        very thick,
                        dashed
                    ]{sqrt(1+x^2)-1};
        \end{axis}
        \fill[red] (5.75,2.1) circle (0.05) node[right]{$\phi_\text{end}$};
        \draw[dashed,red] (5.75,2.1) -- (5.75,0);
    \end{tikzpicture}
    \hspace{1cm}
    \begin{tikzpicture}[scale=0.75, transform shape]
       \begin{axis}[
                    axis lines = center,
                    axis line style = thick,
                    xlabel = \(\frac{\phi}{M}\),
                    ylabel = {\( \frac{V}{m^2 M^2} \)},
                    label style={font=\Large},
                    xtick=\empty,
                    ytick=\empty
                    ]
            \addplot [
                        domain=-1.5:1.5,
                        range=0:2.5,
                        samples=100, 
                        color=black,
                        very thick
                    ]{0.5*x^2};
            \addplot [
                        domain=-1.0:1.5,
                        range=0:2.5,
                        samples=100, 
                        color=gray,
                        very thick,
                        dashed
                    ]{(1-exp(-sqrt(2/3)*x))^2};
        \end{axis}
        \fill[red] (5.75,1.175) circle (0.05) node[right]{$\phi_\text{end}$};
        \draw[dashed,red] (5.75,1.175) -- (5.75,0);
    \end{tikzpicture}
    \caption{Illustrations of symmetric (left panel) and asymmetric (right panel) potentials near a quadratic minimum where $\phi_\text{end}$ indicates the field value where inflation ends and preheating (presumably) begins. While a shallow symmetric potential opens up on both sides of $\phi=0$, the asymmetric potential does so only for $\phi>0$. Hence, the inflaton feels attractive self-interaction at only one side of the potential during the oscillations.}
    \label{fig:even_odd}
\end{figure}
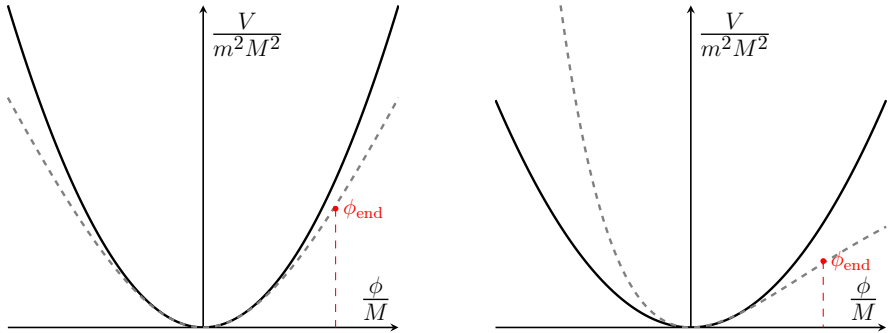\\
\indent In this paper, we examine the possibility of oscillon formation (or, more generally, nonlinear structure formation) in the $\alpha$-attractor E-model \cite{Kallosh:2013hoa,Kallosh:2013yoa} using detailed $3d$ numerical lattice simulations. It is divided into the following sections: in Sec. \ref{sec:small_amp}, we show that oscillon-like solutions can exist in a generic asymmetric polynomial  potential of the form $V(\phi)=\frac{1}{2}\phi^2+A\phi^3+B\phi^4$ using the method of small amplitude analysis in $(1+1)$-$d$. A potential of this form arises from the small amplitude limit of the E-model $\alpha$-attractor  \cite{Kallosh:2013yoa}. In Sec. \ref{sec:Floquet}, Floquet analysis is used to generate the instability chart for inflaton perturbations during the preheating phase which can be used to demonstrate the existence of a broad-band resonance region followed by multiple narrow ones. In Sec. \ref{sec:lattice} we carry out $3d$ lattice simulations of the E-model potential and confirm the formation of nonlinear structures, which can be interpreted as oscillons. By performing a volume integral of the overdensities above a certain threshold, we demonstrate that oscillons can be copiously produced, constituting over $40\%$ of the total energy, for values of $\alpha$ in the relevant range. Finally, in Sec. \ref{sec:discussion} the consequences of the smallness of $\alpha$ is discussed -- particularly in relation to the predicted levels of tensor perturbations and the running of the scalar spectral tilt. There we also comment on the effects of long-term gravitational clustering and some simulation specific issues.\\ 
\indent In this work, we adopt the mostly negative convention for the Friedmann-Lema\^{i}tre-Robertson-Walker (FLRW) metric such that $\dd s^2=\dd t^2-a^2(t)\delta_{ij}\dd x^i\dd x^j$, ignoring the effects of metric perturbations. We also use natural units where $c=\hbar=1$ and the reduced Planck mass is denoted as $\left( 8\pi G \right)^{-1}=M^2_\text{pl}$ which has a numerical value of $2.44\times 10^{18}\:\text{GeV}$.

\section{Small amplitude analysis and existence of oscillon-like cores}\label{sec:small_amp}
\begin{figure}
    \centering
    \includegraphics[scale=0.5]{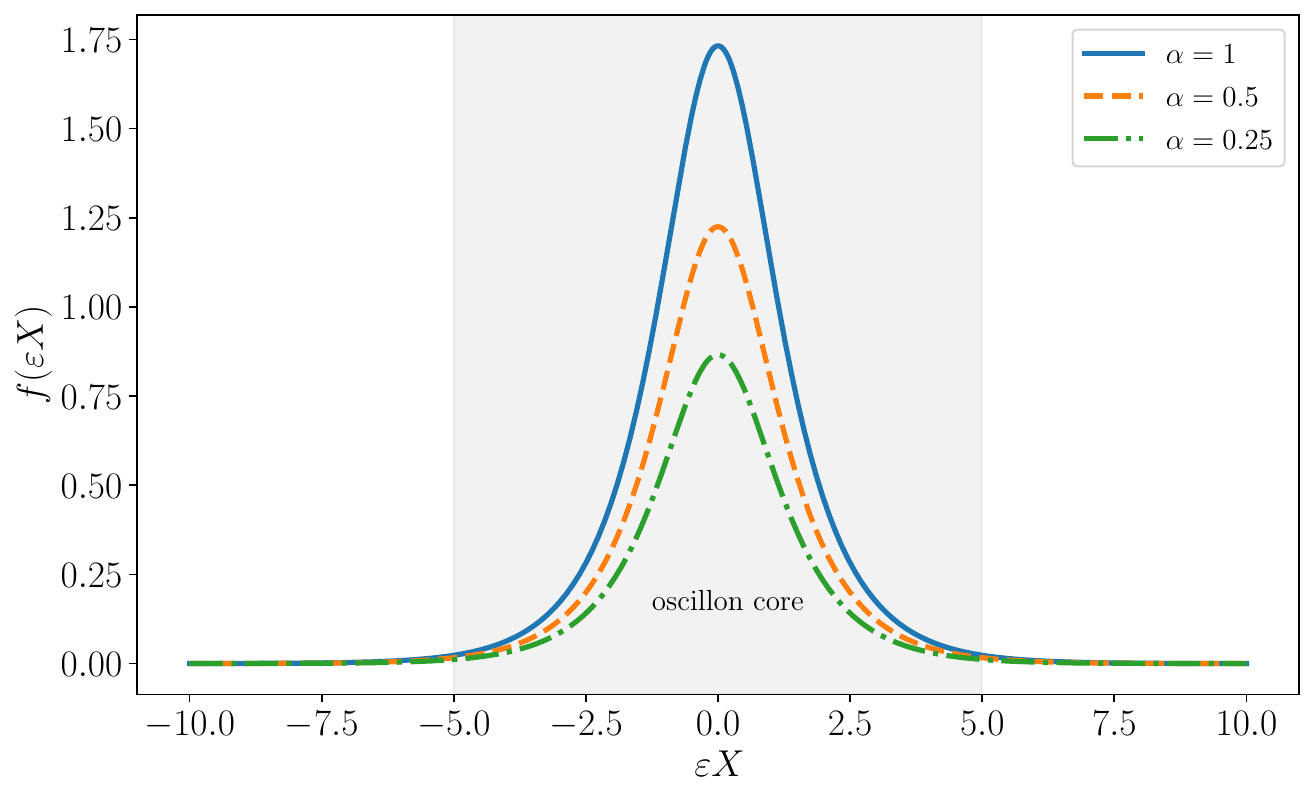}
    \caption{Oscillon core profile obtained from the $4^\text{th}$-order expansion of the E-model potential for different values of $\lambda(\alpha)$.}
    \label{fig:profile2}
\end{figure}

Even though oscillons are nonlinear structures and should be studied in full generality using non-perturbative techniques and numerical simulations, a great deal of information about them can be obtained by using the \textit{small amplitude analysis} \cite{Fodor:2008es,Fodor:2008du,Fodor:2009xw,Fodor:2019ftc,Amin:2010dc,Amin:2010jq,Amin:2010xe}. More importantly, we can use this technique to confirm whether or not a given scalar potential can admit solutions where the field configurations  exhibit oscillon-like cores, which correspond  to  solutions of the nonlinear Klein-Gordon equation with centrally peaked profile oscillating in time. However, since the small amplitude analysis relies upon an expansion in the powers of the scalar field, it is only limited to cases where the amplitude of the scalar field oscillations is much smaller compared to some mass scale (usually the Planck scale $M_\text{pl}$). Moreover, oscillons relevant in the field of cosmology are usually ones with large amplitude. Hence, the small amplitude analysis should only serve as a first step at ascertaining their existence.\\
\indent We consider the $\alpha$-attractor E-model potential with $n=1$ \cite{Kallosh:2013hoa,Kallosh:2013yoa}. For $n>2$, the leading order behavior of the minimum ceases to be quadratic. The E-model potential can be written as
\begin{equation}\label{eq:Emodel}
    V(\phi)=V_0\left[ 1-\exp\left( -\lambda(\alpha)\frac{\phi}{M_\text{pl}} \right) \right]^2
\end{equation}
where $\lambda(\alpha)=\sqrt{2/3\alpha}$. Since Eq. \eqref{eq:Emodel} is an asymmetric potential, we can Taylor expand it around the minimum to derive a polynomial potential that will be easier to study with the small amplitude analysis. Expanding it up to $\mathcal{O}\left( \phi^4 \right)$, we have
\begin{equation}\label{eq:Emodel_taylor}
   U(\phi) \equiv \frac{V(\phi)}{m^2 M_\text{pl}^2}\approx \frac{1}{2}\left( \frac{\phi}{M_\text{pl}} \right)^2-\underbrace{\frac{\lambda}{2}\left( \frac{\phi}{M_\text{pl}} \right)^3}_{\text{\tiny asymmetric}}+\frac{7\lambda^2}{24}\left( \frac{\phi}{M_\text{pl}} \right)^4
\end{equation}
Considering a $(3+1)$-$d$ FLRW metric of the form $\dd s^2=\dd t^2-a^2(t)\delta_{ij}\dd x^i\dd x^j$, we have the following action
\begin{equation}
    S=\int\dd^4 x\sqrt{-g}\left[\frac{M^2_\text{pl}}{2}R+ \frac{1}{2}\partial_\mu\phi\partial^\mu\phi-V(\phi) \right]
\end{equation}
where $\sqrt{-g}$ is the metric determinant and $R$ is the Ricci scalar. The following scalings are performed which transform the variables $(t,x,\phi)$ into dimensionless quantities: $\tilde{t} = mt$, $\tilde{x}=mx$ and $\tilde{\phi}=\phi/M_\text{pl}$. With these redefinitions, and restricting ourselves to $(1+1)$-$d$ in a non-expanding universe, the following equation of motion can be derived
\begin{equation}\label{eq:KG}
    \partial_{\tilde{t}}^2\tilde{\phi}-\partial_{\tilde{x}}^2\tilde{\phi}+\partial_{\tilde{\phi}} U=0
\end{equation}
 Equation \eqref{eq:KG} can be solved order-by-order using a perturbative expansion of $\tilde{\phi}$ and frequency $\omega$ in terms of an expansion parameter $\varepsilon$. To that end, these variables can be written down as
\begin{align}
    \tilde{\phi}&=\sum_{n=1}^{\infty}\varepsilon^n\tilde{\phi}_{(n)}\\
    \omega^{2}(\varepsilon)&=1+\sum_{n=1}^{\infty}\varepsilon^n \omega_{(n)}
\end{align}
Furthermore, we rescale the $(\tilde{t},\tilde{x})$ variables to make them $\varepsilon$-dependent. This is carried out via $y=\varepsilon \tilde{x}$ and $\tau=\omega(\varepsilon) \tilde{t}$. With these, we have the following
\begin{equation}
    -\omega^2\partial_\tau^2\tilde{\phi}+\varepsilon^2\partial_y^2\tilde{\phi}=\tilde{\phi}-\frac{3\lambda}{2}\tilde{\phi}^2+\frac{7\lambda^2}{6}\tilde{\phi}^3 
\end{equation}
In the following steps, we work up to $\mathcal{O}(\varepsilon^3)$ in the expansion and we will show that the first correction to the frequency $\omega$ is obtained precisely at this order. Unlike \cite{Amin:2010jq}, the even powers of $\tilde{\phi}$ will not vanish since $\partial_{\tilde{\phi}}U$ contains a $\tilde{\phi}^2$ term. The aim is to be able to express the evolution of each $\tilde{\phi}_{(n)}$ in the form
\begin{equation}
    \partial_{\tau}^2\tilde{\phi}_{(n)} +\tilde{\phi}_{(n)} = \mathcal{F}_{(n)}\bigg[\tilde{\phi}_{(n-1)},\partial_{\tau}\tilde{\phi}_{(n-1)},\partial_{\tilde{\phi}}U(\tilde{\phi}_{(n-1)})\bigg]
\end{equation}
where $\mathcal{F}_n$ is a forcing term that is a function of the field variable and its derivatives in the previous order in the expansion. Since at each order $n$ we are seeking a differential equation of the form of a forced harmonic oscillator, we shall assume that the solutions do not have very strong spatial dependence. It is trivial to show that, at $\mathcal{O}(\varepsilon)$, the first order field is a simple harmonic oscillator $\partial_{\tau}^2\tilde{\phi}_{(1)} + \tilde{\phi}_{(1)}=0$, with a solution that behaves as $\tilde{\phi}_{(1)}(\tau,y)=f(y)\cos\tau+\mathcal{O}(\varepsilon^2)$.\\
\indent At $\mathcal{O}(\varepsilon^2)$, we have the following
\begin{equation}\label{eq:phi2}
    \omega_{(1)}\partial_\tau^2\tilde{\phi}_{(1)}+\partial^2_{\tau}\tilde{\phi}_{(2)}+\tilde{\phi}_{(2)}-\frac{3\lambda}{2}\tilde{\phi}^2_{(1)}=0
\end{equation}
The term $\partial_{\tau}^2\tilde{\phi}_{(1)}\sim\cos\tau$ is a resonance term which will produce a term in the solution which grows as $\tau$. Since we are interested in bounded solutions, we set $\omega_{(1)}=0$. Solving Eq. \eqref{eq:phi2} using the initial conditions $\tilde{\phi}_{(2)}(0,y)=\partial_{\tau}\tilde{\phi}_{(2)}(\tau,y)|_\tau=0$, we have 
\begin{equation}
    \tilde{\phi}_{(2)}(\tau,y)=\frac{\lambda}{4}f^2(y)\big( 3-2\cos\tau-\cos2\tau \big)
\end{equation}
Similarly, at $\mathcal{O}(\varepsilon^3)$, the evolution of $\tilde{\phi}$ is given by
\begin{equation}\label{eq:phi3}
    \partial_y^2\tilde{\phi}_{(1)}-\omega_{(2)}\partial^2_\tau\tilde{\phi}_{(1)}-\partial^2_\tau\tilde{\phi}_{(3)}-\frac{7\lambda^2}{6}\tilde{\phi}^3_{(1)}+3\lambda\tilde{\phi}_{(1)}\tilde{\phi}_{(2)}-\tilde{\phi}_{(3)}=0
\end{equation}
After simplifying Eq. \eqref{eq:phi3}, there also turns out to be a resonance term that needs to be set to zero to obtain bounded solutions. This results in
\begin{equation}\label{eq:profile}
    \frac{\partial^2f(y)}{\partial y^2}+\omega_{(2)}f(y)+\lambda^2f^3(y)=0
\end{equation}
To obtain solutions which are localized and decay to zero at spatial infinity, one requires the condition $\omega_{(2)}<0$. Also, deriving the first integral of motion (conserved energy)\footnote{The first integral of motion can be shown to be the following $$E=\frac{1}{2}(\partial_y f)^2-\frac{a^2}{2}f^2+\frac{\lambda^2}{4}f^4$$obtained by multiplying Eq. \eqref{eq:profile} by $\partial_\tau\varphi$ and integrating by parts.} and imposing conditions of localized solutions, we obtain 
\begin{equation}
a^2\equiv\omega_{(2)}=\frac{\lambda^2}{2}f^2_0 
\end{equation}
Moreover, integration of the conserved energy equation yields the oscillon core profile (shown in Fig. (\ref{fig:profile2}))
\begin{equation}
f(y)=\sqrt{2}\frac{a}{\lambda(\alpha)}\left[ 1-\tanh^2(-ay) \right]^{1/2}=f_0\text{sech}\left( \frac{f_0}{\sqrt{3\alpha}}y \right)
\end{equation}
\\
Hence, up to $\mathcal{O}(\epsilon^2)$, the solution can be written as
\begin{equation}
\tilde{\phi}_\text{osc}(\tau,y)=\varepsilon f(y)\cos\tau+\varepsilon^2\frac{\lambda}{4}f^2(y)\big( 3-2\cos\tau-\cos2\tau \big)+\mathcal{O}(\varepsilon^3)
\end{equation}
with $y=\varepsilon \tilde{x}$ and $\tau=\sqrt{1-\varepsilon^2 a^2}\tilde{t}$. Since $\varepsilon$ is a small expansion parameter, it can be chosen such that $\varepsilon\equiv \Phi_0/M_\text{pl}$. As a result, we can write
\begin{equation}
    \phi_\text{osc}(\tau,y)=\Phi(y)\left[ \cos\tau+\frac{\lambda}{4}\frac{\Phi(y)}{M_\text{pl}}\left( 3-2\cos\tau-\cos2\tau \right) \right]+\mathcal{O}(\varepsilon^3)
\end{equation}
where $\varepsilon M_\text{pl}f(y)=\Phi(y)$. Hence, the solution $\varphi_\text{osc}$ describes spatially localized profiles which display temporal variations with combinations of the frequency $\omega=\sqrt{1-\varepsilon^2 a^2}m\approx\left( 1-\varepsilon^2\frac{\lambda^2 f^2_0}{4} \right)m$, indicating towards the possibility of oscillon formation in such an asymmetric potential. However, we must remember that this is only an approximation and one needs to verify the existence of such nonlinear objects through lattice simulations which is the goal of the succeeding sections. Furthermore, these calculations only considered one spatial dimension and neglected the effects of expansion. For instance, it can be shown that in $d\geq 2$, the energy associated with the oscillon solution decays due to the presence of a friction term \cite{Amin:2010dc}
\begin{equation}
    \partial_y E=-\frac{d-1}{y}\left( \partial_y f \right)^2
\end{equation}
We note that for a general potential of the form $V(\phi)=\frac{1}{2}\phi^2+A\phi^3+B\phi^4$, solutions admitting an oscillon profile may only appear for a range of values in the parameter space of $A$ and $B$. One can show that, in general
\begin{equation}
    f(y)=\sqrt{\frac{16}{3\Gamma(A,B)}}\text{sech}\left(y\right)
\end{equation}
for $a=1$, such that
\begin{equation}
    \Gamma(A,B)=20A^2-8B
\end{equation}
\begin{figure}
    \centering
    \includegraphics[scale=0.5]{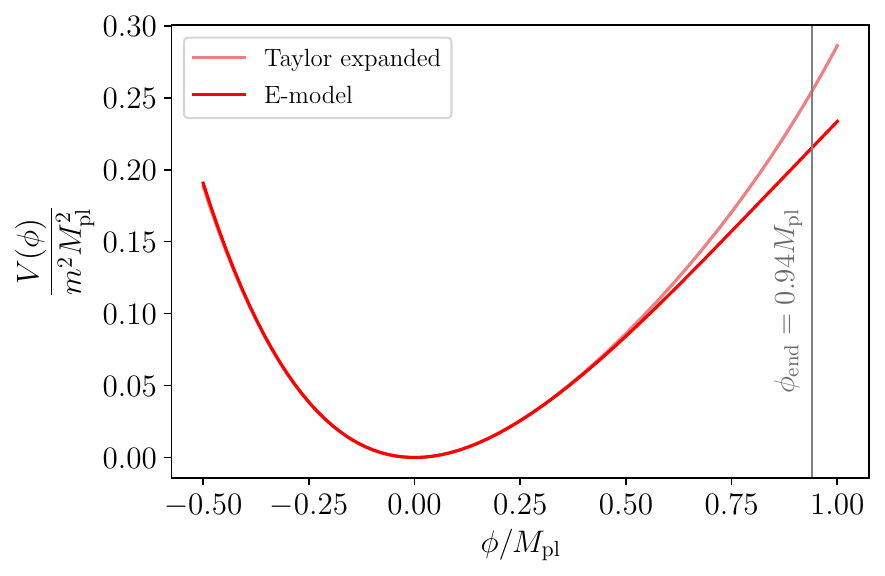}
    \includegraphics[scale=0.5]{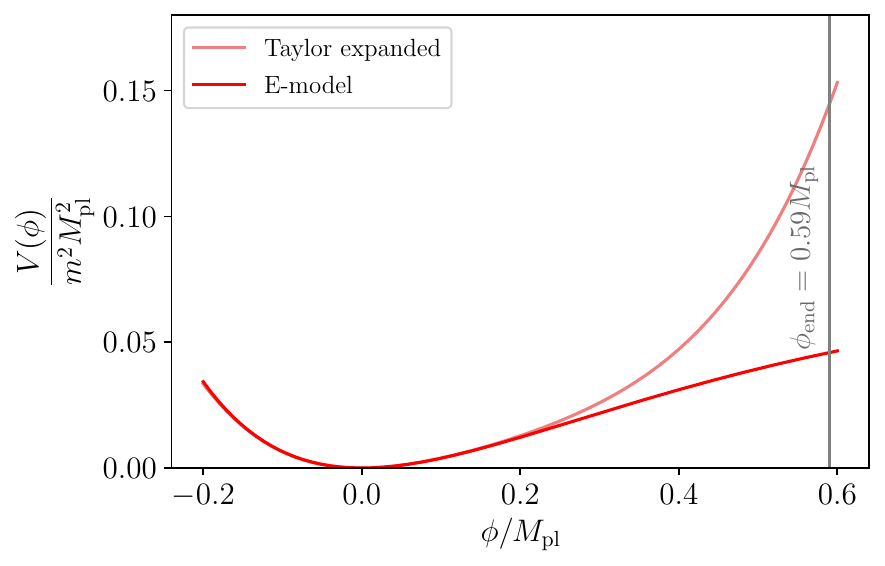}
    \caption{Comparisons of the E-model and $4^\text{th}$-order polynomial potentials for $\alpha=1$ (left panel) and $0.1$ (right panel). The gray vertical lines correspond to the field values where inflation ends and, hence, preheating commences. This figure aims to show that the Taylor expansion should really be used to model the E-model potential when $\alpha\sim\mathcal{O}(1)$.}
    \label{fig:potentials_plots}
\end{figure}\\
For a real solution for $f(y)$, we require that $\Gamma(A,B)>0$, which restricts the possible values for $A$ and $B$. Although we have shown that oscillon-like solutions exist for an asymmetric potential, it does not necessarily mean that the inflaton condensate fragments and forms oscillons during preheating. We further note that Eq. \eqref{eq:Emodel_taylor} does not perfectly capture details of the E-model, especially in relation to its plateau region for $\phi>0$. If we use the field value at the end of inflation as the initial condition for the preheating stage, it can be observed that the values of the potentials are within agreeable limits when $\alpha\sim\mathcal{O}(1)$ and the difference in the values of $V(\phi)$ increases sharply for lower values of $\alpha$. This is shown in Fig. (\ref{fig:potentials_plots}). In what follows, we perform a detailed $3d$ lattice study on the E-model in pursuit of oscillon formation. For the reasons stated above, we will not use the Taylor expanded potential in the lattice simulations. 

\section{Instability bands of the E-model}\label{sec:Floquet}
Parametric resonance plays a crucial role in governing the growth of perturbations during inflationary preheating \cite{Amin:2014eta,Lozanov:2019jxc}. As a result, it is important to identify the existence of instability bands in the parameter space of comoving modes $k$ and  the field amplitude and to study the trajectory  of modes as they pass  through the instability bands due to their  time evolution. A linear stability analysis of inflaton perturbations $\delta\phi_{\bm{k}}$ can be performed using Floquet theory. In the next subsections, Floquet theory is introduced and applied to the E-model to identify the instability bands of the model. 
\subsection{Linear stability analysis and Floquet theory}
Before applying Floquet theory, we can study the inflaton fluctuations at the linear order in perturbation theory and search for instability bands. This analysis can be carried out  with or without the inclusion of metric perturbations. Specializing  to the case were metric perturbations are absent, we consider the inflaton Klein-Gordon equation $\ddot{\phi}+3H\dot{\phi}-\frac{\grad^2}{a^2}\phi+\partial_\phi V=0$ and linearize it around scalar fluctuations $\phi(t,\bm{x})=\phi(t)+\delta\phi(t,\bm{x})$, obtaining the following equation for the perturbations
\begin{equation}
    \delta\ddot{\phi}+3H\delta\dot{\phi}+\partial_{\phi\phi}V(\phi)\delta\phi-\frac{\grad^2}{a^2}\delta\phi=0
\end{equation}
or, in Fourier space
\begin{equation}\label{eq:phi_pert}
    \delta\ddot{\phi}_{\bm{k}}+3H\delta\dot{\phi}_{\bm{k}}+\partial_{\phi\phi}V(\phi)\delta\phi_{\bm{k}}+\frac{k^2}{a^2}\delta\phi_{\bm{k}}=0
\end{equation}
with Bunch-Davies vacuum conditions $\delta\phi_{\bm{k}}=1/( a\sqrt{2k} )$ and $\delta\dot{\phi}_{\bm{k}}=-i\left( 
k/a \right)\delta\phi_{\bm{k}}$. If we consider, to leading order, that the minimum of $V(\phi)$ is quadratic, then the inflaton oscillates about it such that $\phi(t)=\phi_0(t) \cos (mt)$. Ignoring cosmic expansion for the moment (for which $\phi_0$ becomes  time independent) and setting $a=1$, Eq. \eqref{eq:phi_pert} takes the form of a parametric oscillator\footnote{If we consider that the typical time period of oscillations is much less than the Hubble timescale, \textit{i.e.} $T=2\pi/\omega\ll H^{-1}$, then the expansion term can be ignored since $\omega_{\bm{k}}(t)\gg H(t)$. This is usually true after a few oscillations when $H \propto 1/t$ drops well below the inflaton mass.}
\begin{equation}\label{eq:phi_pert_k}
    \delta\ddot{\phi}_{\bm{k}}+\left[ k^2+\partial_{\phi\phi}V(\phi(t)) \right]\delta\phi_{\bm{k}}=0
\end{equation}
with the term $k^2+\partial_{\phi\phi}V$ serving as the time-dependent frequency $\omega_{\bm{k}}^2(t)$. When metric perturbations were included, the mode equation would take the form
\begin{equation}\label{eq:phi_pert_metric}
    \delta\ddot{\phi}_{\bm{k}}+\omega^2_{\bm{k}}(t)\delta\phi_{\bm{k}}=\underbrace{4\delta\dot{\phi}_{\bm{k}}\dot{\Psi}-2\Psi\partial_{\phi}V}_{\text{metric perturbations}}
\end{equation}
where $\Psi$ is the Bardeen potential. Second order differential equations of these forms are also known as the \textit{Hill's differential equation} \cite{nwm,abramowitz+stegun,822801} and, according to Floquet theorem, they admit solutions of the form
\begin{equation}
    \delta\phi_{\bm{k}}(t)=e^{\mu_{\bm{k}}t}\mathcal{P}_{(+)}(t)+e^{-\mu_{\bm{k}}t}\mathcal{P}_{(-)}(t)
\end{equation}
The quantities $\mu_{\bm{k}}$ in the exponents are called Floquet exponents and they describe how the modes grow or decay with time. The functions $\mathcal{P}_{(\pm)}(t)$ are periodic functions with the same period of oscillation as the background field. The $\mu_{\bm{k}}$ are complex in general and those modes with $\mathfrak{R}(\mu_{\bm{k}})>0$ correspond to growing solutions.\footnote{This can be seen as the result of violation of the adiabatic condition $|\dot{\omega}/\omega^2| < 1$. As a consequence, the perturbations grow as $\delta\phi_{\bm{k}}\sim e^{\mu_{\bm{k}}t}$ instead of oscillating as $\delta\phi_{\bm{k}}\sim e^{\pm i\int\omega_{\bm{k}}\dd t}$.} For purely imaginary exponents, the modes are oscillatory. The utility of Floquet theory lies in the fact that one can map such `zones of instability' in the form of a Floquet chart. Such a chart comprises of bands of instability in the form of broad and narrow resonance bands.\\
\indent Apart from a few simple cases, the Floquet exponents need to be computed numerically. However, there exists a relatively straightforward approach using which the Floquet exponents can be numerically calculated \cite{Amin:2014eta,Lozanov:2019jxc,Johnson:2008se}. We start by expressing Eq. \eqref{eq:phi_pert_k} as a system of coupled differential equations in matrix form
\begin{equation}
    \delta\dot{\bm{\Phi}}(t)=\mathcal{U}(t)\delta\bm{\Phi}(t)
\end{equation}
where $\delta\bm{\Phi}(t)=\left[ \delta\phi_{\bm{k}},\delta\pi_{\bm{k}} \right]^\text{T}$ and $\delta\pi_{\bm{k}}=\delta\dot{\phi}_{\bm{k}}$. The matrix $\mathcal{U}(t)$ is given by
\begin{equation}
    \mathcal{U}(t)=\mqty( 0 & 1 \\ -k^2-\partial_{\phi\phi}V & 0 )
\end{equation}
Next, the time period $T$ of the oscillation is determined. If the potential is symmetric, this is as simple as performing the integral
\begin{equation}
    T\left( \phi_\text{\tiny{in}} \right)=2\int_{0}^{\phi_{\text{\tiny{in}}}}\frac{\dd\phi}{\sqrt{2V(\phi_{\text{\tiny{in}}})-2V(\phi)}}\equiv2T_{(1/2)}
\end{equation}
where $T_{(1/2)}$ is the period of a half oscillation. However, if the potential is asymmetric, as in the present case, the time period cannot be calculated using the expression above. For odd potentials, one can show that $T$ takes the following form
\begin{align}
    T&=T^{(+)}+T^{(-)} \nonumber\\
     &=\int_{0}^{\phi_{\text{in}}}\frac{\dd\phi}{\sqrt{2V(\phi_{\text{\tiny{in}}})-2V(\phi)}}+\int_{\phi_\text{f}}^{0}\frac{\dd\phi}{\sqrt{2V(\phi_{\text{\tiny{f}}})-2V(\phi)}}
\end{align}
where $\phi_\text{\tiny f}$ is the amplitude to which the field rises on the $\phi<0$ side of the potential that can be determined using energy considerations,  namely $V(\phi_{\rm \tiny{f}}) = V(\phi_{\rm \tiny{in}})$. The time period can also be calculated using the solution of the background evolution. Now, using a set of orthogonal initial conditions $\{ \delta\phi^{(1)}_{\bm{k}}(0)=1,\delta\pi^{(1)}_{\bm{k}}(0)=0 \}$ and $\{ \delta\phi^{(2)}_{\bm{k}}(0)=0,\delta\pi^{(2)}_{\bm{k}}(0)=1 \}$, Eq. \eqref{eq:phi_pert_k} is evolved from $t=0$ to $T$. Then, the Floquet exponents are calculated by
\begin{equation}\label{eq:floquet}
    \mathfrak{R}\left( \mu_{\bm{k}}^{(\pm)} \right)=\frac{1}{T}\ln\left[ \frac{1}{2}\left( \delta\phi_{\bm{k}}^{(1)}+\delta\pi_{\bm{k}}^{(2)}\right)\pm\frac{1}{2}\sqrt{\left( \delta\phi_{\bm{k}}^{(1)}-\delta\pi_{\bm{k}}^{(2)} \right)^2+4\delta\phi_{\bm{k}}^{(2)}\delta\pi_{\bm{k}}^{(1)}}  \right]
\end{equation}
such that
\begin{equation}
    \mathfrak{R}\left( \mu_{\bm{k}} \right)=\max\left[ \mathfrak{R}\left( \mu_{\bm{k}}^{(\pm)} \right) \right]
\end{equation}
In Eq. \eqref{eq:floquet}, all the quantities are evaluated at $t=T$. Hence, the calculation of the instability bands for parametric resonance essentially involves evolving these mode functions using orthogonal initial conditions and searching for the existence of $\mathfrak{R}\left( \mu_{\bm{k}} \right)>0$. When evaluated for an entire grid of $\left( k,\phi_\text{in} \right)$, the instability regions of the parametric oscillator can be mapped out, which typically display banded structures.

\subsection{Results for the E-model}
We now consider parametric resonance occurring in the $\alpha$-attractor E-model. The potential can be written down as follows
\begin{equation}
    V(\phi)=\frac{3}{4}\alpha m^2 M^2_\text{pl}\left( 1-e^{-\sqrt{\frac{2}{3\alpha}}\frac{\phi}{M_{\text{pl}}}} \right)^2
\end{equation}
\begin{figure}
    \centering
    \includegraphics[scale=0.45]{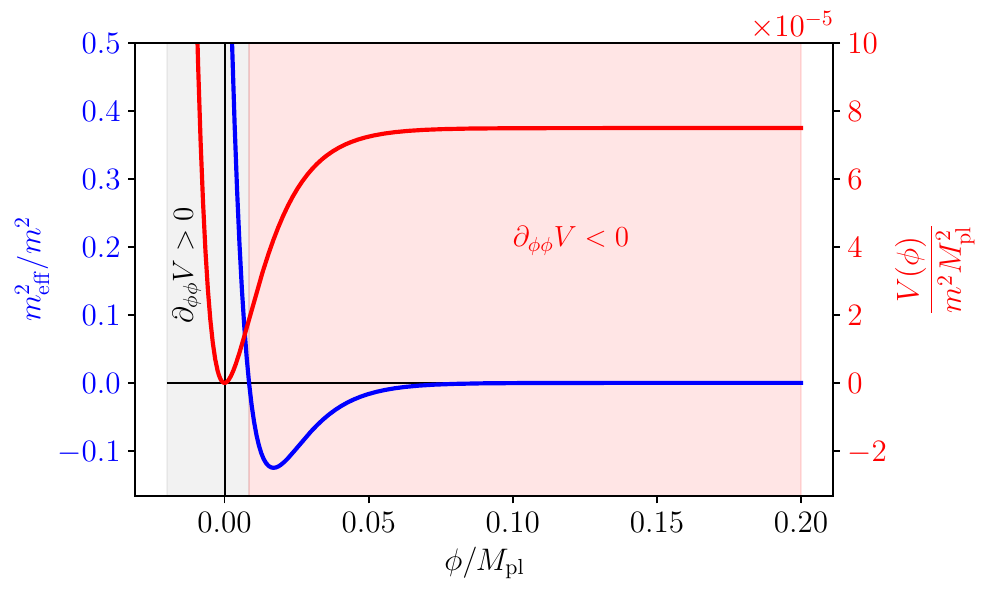}
    \includegraphics[scale=0.45]{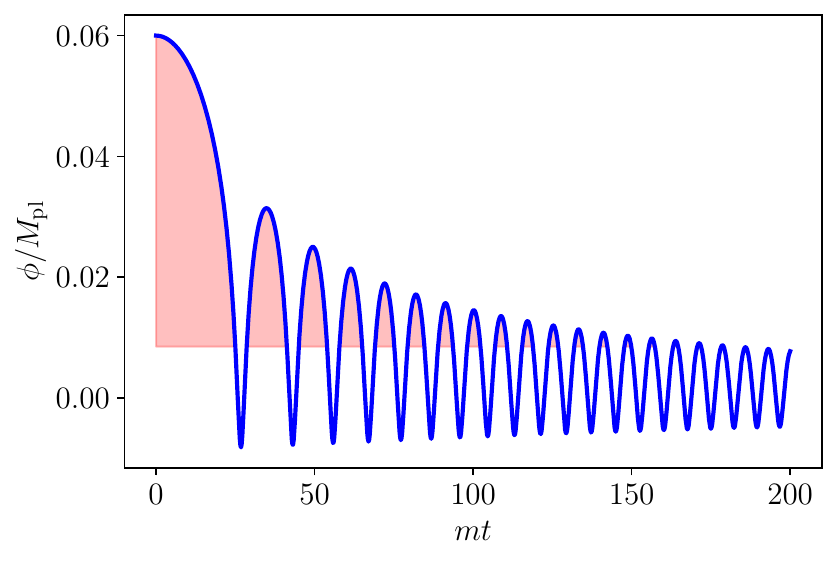}
    \caption{\textit{Left panel}: The effective mass squared $m^2_\text{eff}$ (blue) and the E-model potential (red) are shown for $\alpha=10^{-4}$. The red shaded region indicates the field values for which $m^2_\text{eff}<0$. In the plateau region, the effective mass is still negative but asymptotically approaches zero. \textit{Right panel}: The homogeneous background evolution of the field is shown where the red shaded portions indicate the regions where the field oscillations enter the tachyonic instability regimes. As the field amplitude decays due to expansion, the inflaton is unable to access the tachyonic instability regime after a fixed time has passed.}
    \label{fig:meff}
\end{figure}\\
The overall constant being $V_0=\frac{3}{4}\alpha m^2 M^2_\text{pl}$ reflects the fact that the potential, at leading order, resembles $\frac{m^2\phi^2}{2}$. Since the background and mode evolution equations can be made dimensionless, the only relevant parameter in the potential is $\alpha$. It is important to note that inflaton potentials with a plateau generally exhibit tachyonic instability where the effective mass squared $m^2_\text{eff}$ becomes negative \cite{Felder:2000hj,Felder:2001kt}. This is also true for the E-model potential which possesses a plateau at large $\phi$. For the E-model, it can be shown that $m^2_\text{eff}\equiv\partial_{\phi\phi}V<0$ when $\phi>\sqrt{\frac{3\alpha}{2}}0.693M_\text{pl}$. This is shown in Fig. (\ref{fig:meff}). In the left panel we have plots of $m_\text{eff}^2$ (solid blue) and $V(\phi)$ (solid red) where we see that $m^2_\text{eff}<0$ extends throughout the plateau, becoming prominent near the plateau's edge.\footnote{The effective mass for the E-model is $$ \frac{m_\text{eff}^2}{m^2}=2e^{-2y}-e^{-y}\;\;\;\text{where}\;\;\; y=\sqrt{\frac{2}{3\alpha}}\phi $$For asymptotically large field values, $m^2_\text{eff}\simeq -e^{-y}m^2$. Hence the effective mass stays negative throughout, although very nearly zero for large $\phi$, until $\phi=\sqrt{\frac{3\alpha}{2}}0.693M_\text{pl}$.} The homogeneous field evolution is shown in the right panel. There we observe that during oscillations, $\phi$ repeatedly enters and exits the tachyonic instability region for a few cycles, before cosmic expansion dampens out the amplitude. A detailed study of tachyonic preheating in plateau inflation can be found in Ref. \cite{Tomberg:2021bll}.
\begin{figure}[h]
    \centering
    \includegraphics[scale=0.65]{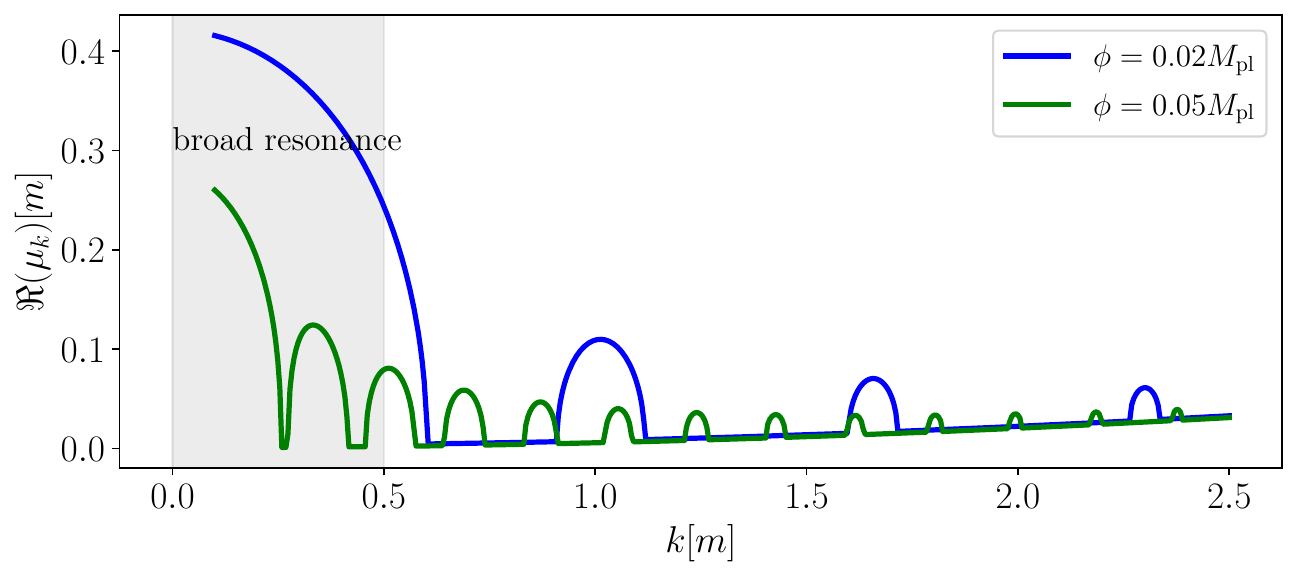}
    \caption{Variation of the Floquet exponent as a function of $k$ at two different values of the field amplitude. The gray shaded region indicates the broad resonance regime which will be pronounced for $0.02\lesssim\phi/M_\text{pl}\lesssim0.03$ for $\alpha=10^{-4}$.}
    \label{fig:mu_k_field}
\end{figure}
\begin{figure}[h]
    \centering
    \includegraphics[scale=0.5]{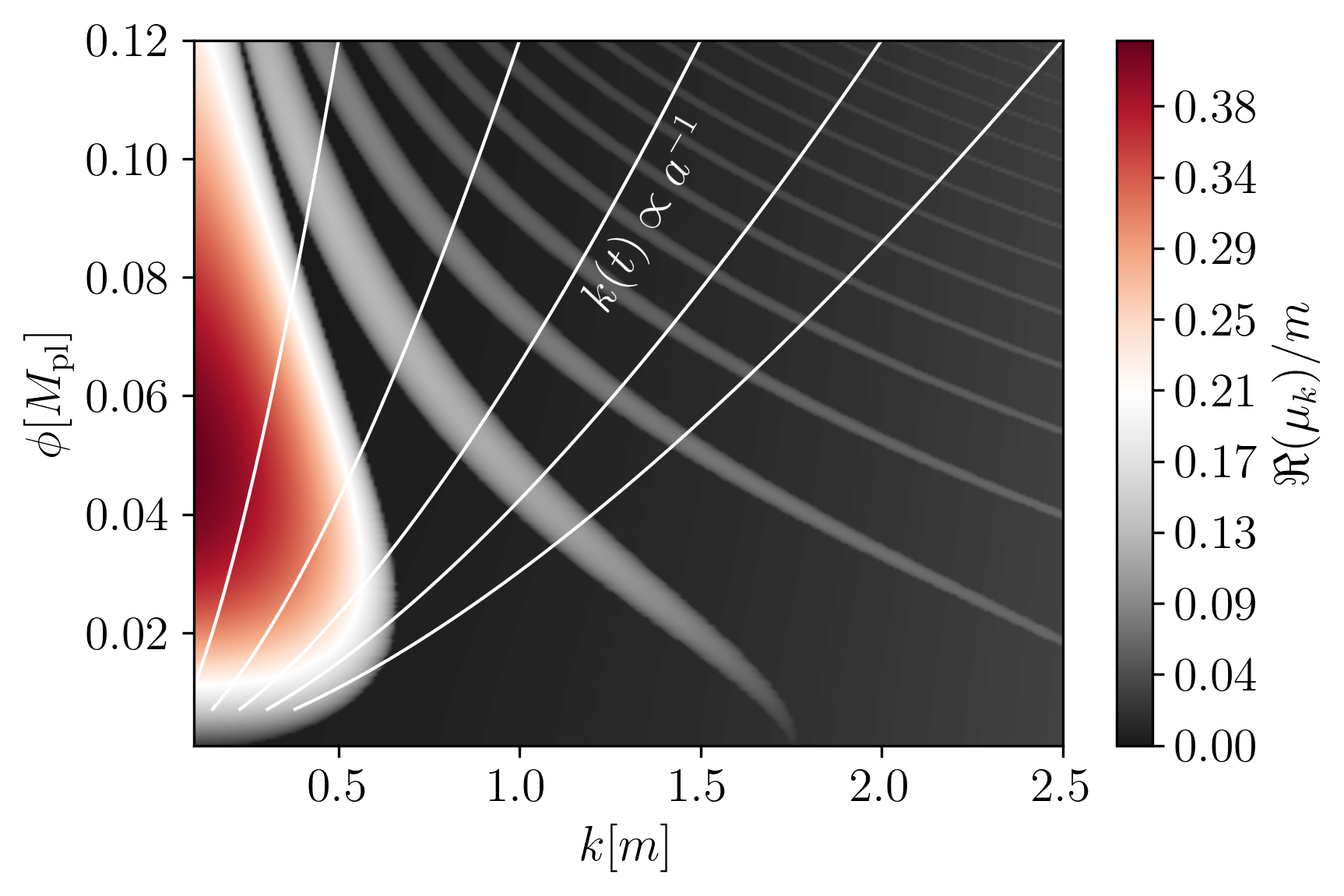}
    \includegraphics[scale=0.5]{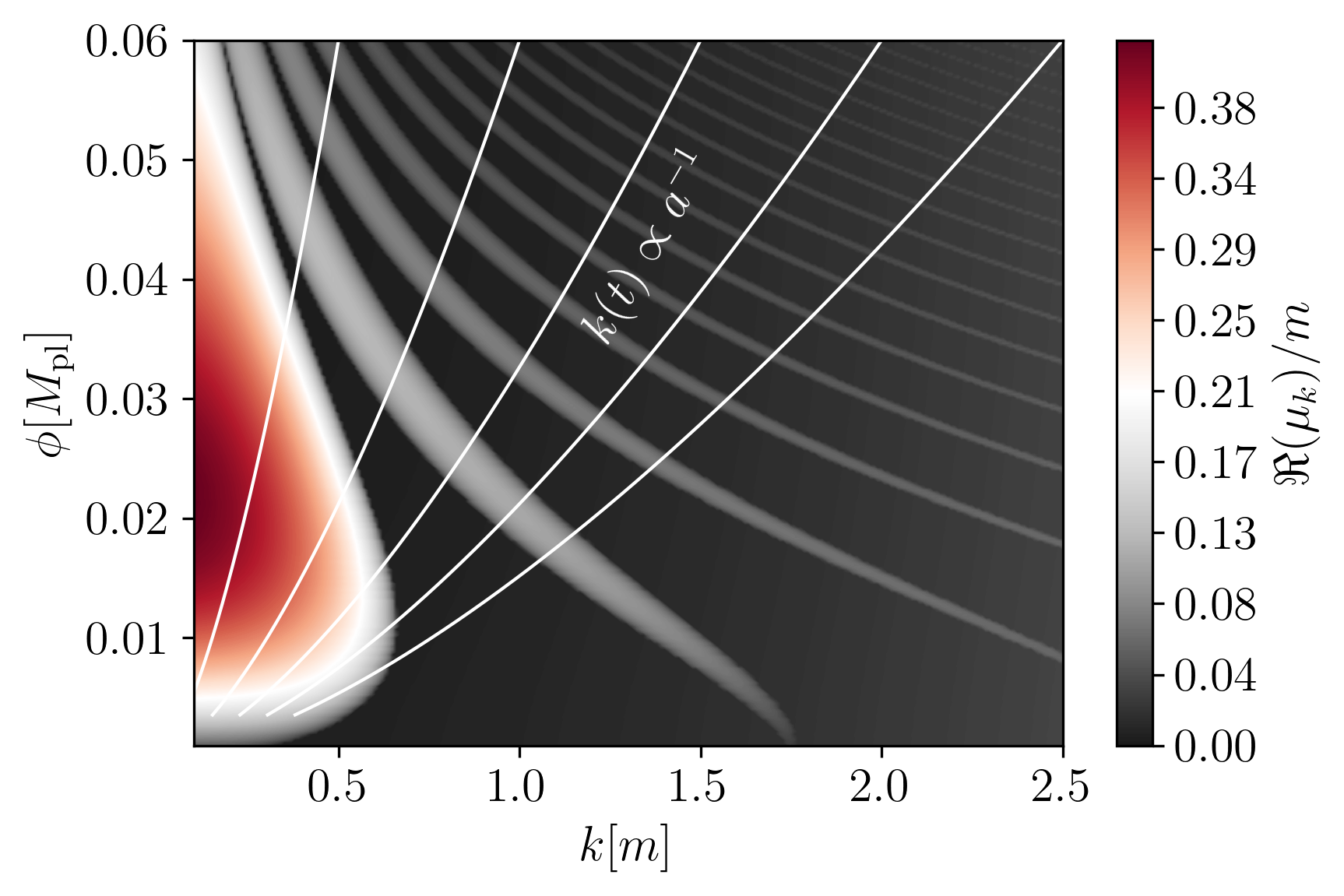}
    \caption{Floquet charts for the E-model with $\alpha=5\times 10^{-4}$ (left panel) and $\alpha=10^{-4}$ (right panel) with the colorbar denoting the real part of the Floquet exponent $\frac{\mathfrak{R}(\mu_k)}{m}$. The charts show the existence of a broad instability band for $k\lesssim0.5m$, where $\left[\frac{\mathfrak{R}(\mu_k)}{m}\right]_\text{max}\sim 0.40$, and several narrower ones for larger $k$. The white flow lines show how the physical $k$-modes pass through different resonance bands as they evolve.}
    \label{fig:floquet_charts}
\end{figure}\\
\indent In Fig. (\ref{fig:mu_k_field}) the variation of the Floquet exponents is shown for two different field values while in Fig. (\ref{fig:floquet_charts}), the Floquet charts for the E-model are plotted for $\alpha=5\times 10^{-4}$ and $10^{-4}$.\footnote{For the benefit of the reader, we note that the different $\phi[M_\text{pl}]$ in Fig. (\ref{fig:mu_k_field}) and (\ref{fig:floquet_charts}) refer to the field amplitudes for the $\phi>0$ excursions of the inflaton. This distinction is necessary due to the asymmetric nature of the E-model potential.} They show the existence of a broad instability band for $k\lesssim 0.5 m$ and a series of narrower ones for higher values of $k$. Since the physical wavenumber in an expanding universe is $k/a(t)$, for a given amplitude, the Fourier modes will trace out a path in the instability chart and will pass through one or several of these instability bands. Specifically, if the potential is quadratic near the minimum, one finds $\phi(t)\propto a^{-3/2}(t)\phi_{\text{in}}$ and $k_\text{phys}(t)\propto a^{-1}(t)$. For example, a small $k$-mode can exhibit strong growth as it passes through the broad resonance band as it evolves through time, eventually decreasing as it moves out the band. Likewise, a larger $k$-mode can initially pass through several of the narrow bands before entering the broad resonance band at late times. The evolution of the $k$-modes are shown by the white lines in the instability chart. The crossing of these bands with time can qualitatively explain how the power spectrum $\mathcal{P}_{\delta\phi}(k)$ of the fluctuations should evolve. In particular, one should expect the following
\begin{itemize}
    \item For comoving modes which are $k\lesssim 0.5m$, there should be a sharp enhancement in $\mathcal{P}_{\delta\phi}(k)$ at relatively early times. We use the word `relatively' to emphasize the fact that the modes do not start out in the broad resonance band at $t=0$. However, these are the first modes to experience the strong amplification in the perturbations since the $k_\text{phys}$ corresponding to these comoving modes evolve through to the broad resonance regime the earliest.
    \item Larger comoving $k$-modes can experience slight enhancements at early times due to the passage of the corresponding $k_\text{phys}$ through the narrow resonance bands. At later times, when they enter the broad resonance band, the amplification of these modes will be significantly enhanced and we should expect the peak, which is initially centered at small $k$, to be shifted to larger $k$ values.
\end{itemize}
This will be explored in detail in the next section where lattice simulation results are interpreted.

\section{Lattice results}\label{sec:lattice}
Here we present the results of $3d$ lattice simulations on the E-model for different values of $\alpha$. There are a number of publicly available lattice codes that can be used for studying preheating dynamics -- \textsf{LatticeEasy} \cite{Felder:2000hq}, \textsf{DEFROST} \cite{Frolov:2008hy}, \textsf{HLattice} \cite{Huang:2011gf} and \textsf{PyCool} \cite{Sainio:2012mw} to name a few. We use the publicly available lattice code \textsf{$\mathcal{C}\text{osmo}\mathcal{L}\text{attice}$} \cite{Figueroa:2020rrl,Figueroa:2021yhd} for our $3d$ simulations. In this section, we first discuss numerical simulation setup and then move on to analyzing the simulation results. Finally, we discuss whether or not the nonlinearities observed from the E-model can be interpreted as oscillons.
\subsection{$\mathcal{C}\text{osmo}\mathcal{L}\text{attice}$ setup and parameters}
Lattice simulations work with dimensionless variables.  In terms of the dimensionless variables as $\tilde{t}=mt$, $\tilde{x}=mx$ and $\tilde{\phi}=\frac{\phi}{M_\text{pl}}$, the inhomogeneous inflaton Klein-Gordon equation to be solved takes the form
\begin{equation}\label{eq:inflaton_KG_inhom}
    \ddot{\tilde{\phi}}-\frac{\widetilde{\grad}^2}{a^2}\tilde{\phi}+3\frac{\dot{a}}{a}\dot{\tilde{\phi}}+\partial_{\tilde{\phi}}\widetilde{V}=0
\end{equation}
where the redefined potential now reads
\begin{equation}
    \widetilde{V}(\tilde{\phi})=\frac{3}{4}\alpha\left( 1-e^{-\sqrt{\frac{2}{3\alpha}}\tilde{\phi}} \right)^2
\end{equation}
In order to consistently solve the preheating dynamics, one must also account for the background expansion of space through the Friedmann and Raychaudhuri equations, given by
\begin{align}
    H^2\equiv\left( \frac{\dot{a}}{a} \right)^2&=\frac{1}{3M_\text{pl}^2}\big\langle E_\text{\tiny K}+E_\text{\tiny G}+E_\text{\tiny V} \big\rangle \\
    \frac{\ddot{a}}{a}&=\frac{1}{3M_\text{pl}^2}\big\langle -2E_\text{\tiny K}+E_\text{\tiny V} \big\rangle
\end{align}
where $E_\text{\tiny K}$, $E_\text{\tiny G}$ and $E_\text{\tiny V}$ are the kinetic, gradient and potential energies associated with the inflaton field and $\big\langle \cdot\cdot\cdot \big\rangle$ represents spatial averaging. \textsf{$\mathcal{C}\text{osmo}\mathcal{L}\text{attice}$}, like most other lattice codes, does not take into account the effects of metric perturbations arising from the Bardeen potential $\Psi(t,\bm{x})$, because of which we cannot study the nonlinearities arising from gravitational clustering on longer time scales. As previously mentioned, the only free parameter involved in this model is $\alpha$. The initial field values for the lattice simulations are chosen such that they correspond to the point in field space where inflation ends $(\phi_\text{end})$, which depends on the choice of $\alpha$. For each value of $\alpha$, the initial field values are chosen such that constraints on inflationary observables are satisfied -- namely the CMB normalization $\mathcal{P}_\zeta = 2.1\times 10^{-9}$ at the pivot scale $k_\star = 0.05\:\text{Mpc}^{-1}$, the scalar spectral index $n_s\approx 0.965$ and the scalar-to-tensor ratio $r<0.036$ \cite{Planck:2018jri,Planck:2018vyg,BICEPKeck:2022mhb}. These details are presented in Appendix \ref{sec:inflationary_parameters}.\\
\indent The numerical simulations are performed for the following parameters and corresponding initial conditions for the homogeneous field \\
\begin{table}[h]
    \centering
    \begin{tabular}{@{}lll@{}}
        \toprule
        $\alpha$          & $\phi_\text{in}[M_\text{pl}]$ & $\phi_\text{in}[\text{GeV}]$ \\ \midrule
        $5\times 10^{-4}$ & $0.109$             & $2.65\times10^{17}$  \\
        $10^{-4}$         & $0.058$             & $1.42\times 10^{17}$ \\
        $10^{-5}$         & $0.023$             & $5.57\times 10^{16}$ \\ \bottomrule
    \end{tabular}
    \caption{The different choices of the parameter $\alpha$ along with the corresponding $\phi_\text{in}$.}
    \label{tab:parameters}
\end{table}\\
Table \ref{tab:parameters} summarizes the different choices of $\alpha$ used in the simulations along with the corresponding initial value for the field $\phi_\text{in}$. With an inflationary excursion corresponding to $55\:e$-folds since the pivot scale becomes superhorizon, we find that the mass scale for this model is set to $m\approx 1.27\times 10^{-5} M_\text{pl}$, which does not change irrespective of the value of $\alpha$. The main results of this work were obtained from simulations with lattice size $N=256^3$ and $\tilde{k}_\text{\tiny IR}=0.05$ (here $\tilde{k}_\text{\tiny IR}$ is the minimum infrared cut-off\footnote{The infrared cut-off is defined as $$ \tilde{k}_\text{\tiny IR}=\frac{2\pi}{\tilde{L}} $$where $\tilde{L}$ is comoving length of the cubic lattice. With $\tilde{k}_\text{\tiny IR}=0.05$, the cubic lattices have sides of comoving lengths $\tilde{L}=125.6m^{-1}$.} for the reciprocal lattice) while some results pertaining to longer time evolution, mainly the time evolution of the equation of state (EoS) parameter, were obtained with $N=128^3$. Moreover, the time integration was carried out using the $2^\text{nd}$-order Velocity-Verlet (VV2) algorithm available in \textsf{$\mathcal{C}\text{osmo}\mathcal{L}\text{attice}$}. Higher order VV algorithms can also be used. However, we found no quantitative difference in the outputs of the spatially averaged quantities and, as a result, VV2 is a perfectly reasonable choice for a time integrator in terms of accuracy and speed. We do note that, at least for the E-model, there will be significant numerical differences for lower resolution grids, starting from $N=64^3$, with the solutions displaying \textit{overshooting} at around $\tilde{t}\sim 50$. Hence, we recommend that the smallest grid size with which the simulations are to be  carried out should be $N=128^3$.\\
\indent It should also be of interest to analytically check whether fragmentation of the inflaton condensate occurs for the  chosen parameters. In Refs. \cite{Kim:2017duj,Cotner:2019ykd,Kim:2021ipz}, bounds on the parameters of $\alpha$-attractor T and E-models for inflaton fragmentation were derived. For an asymmetric potential of the form $V(\phi)=\frac{1}{2}m^2\phi^2-A\phi^3$, it was found that
\begin{equation}
    A>\frac{400}{9\sqrt{6\pi}}\left( \frac{m^2}{M_\text{pl}} \right)\left( \frac{0.1}{r_\text{\tiny A}} \right)\approx 5.78\left( \frac{m^2}{M_\text{pl}} \right)\left( \frac{0.1}{r_\text{\tiny A}} \right)
\end{equation}
where $r_\text{\tiny A}\approx 0.1$. This can be used provide bounds on the value of $\alpha$ necessary for fragmentation. One finds that
\begin{equation}\label{eq:alpha_estimate}
    \alpha \lesssim 5\times 10^{-3}\left( \frac{r_\text{\tiny A}}{0.1} \right)^2
\end{equation}
Hence, according to this analytical estimate, we should be able to observe inflaton fragmentation for all the three parameters listed in Table \ref{tab:parameters}.
\subsection{Backreaction from inhomogeneous dynamics}
\begin{figure}
    \centering
    \includegraphics[scale=0.5]{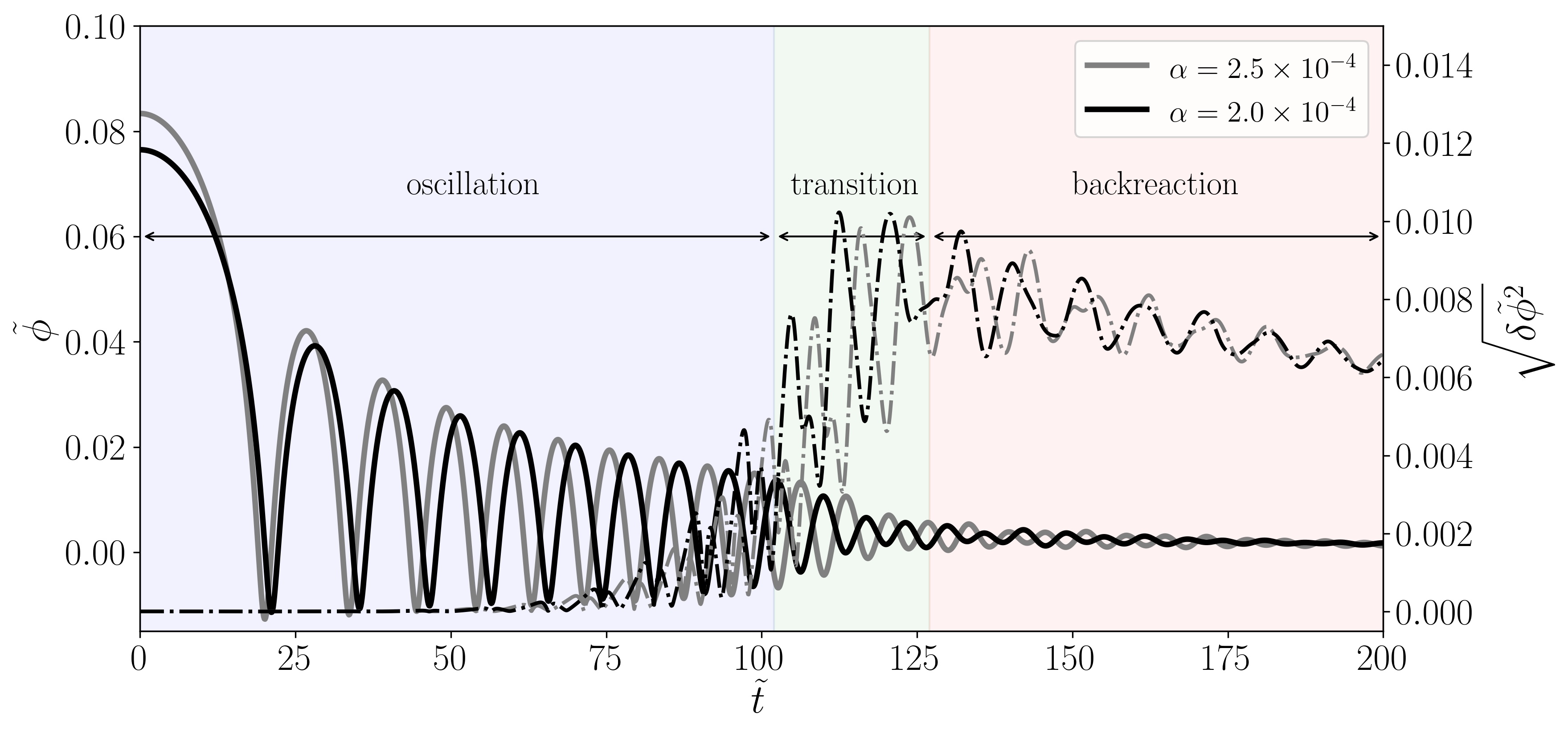}
    \caption{This figure demonstrates that backreaction effects become prominent for $\alpha \leq 2\times 10^{-4}$. The solid lines and dash-dotted lines represent $\tilde{\phi}$ and $\sqrt{\delta\tilde{\phi}^2}$ respectively. The evolution of the field can be roughly divided into three parts -- (i) oscillations, (ii) transition and (iii) backreaction where we observe during the transition period the size of fluctuations becomes of the order of the field value.}
    \label{fig:backreaction_start}
\end{figure}

From Eq. \eqref{eq:inflaton_KG_inhom}, we notice that the presence of the $-\widetilde{\grad}^2\tilde{\phi}$ term can play an important role in the time evolution of the volume averaged field. Although not significant during the slow-roll phase of inflation, the inhomogeneities can, nevertheless, become very important during the preheating phase, leading to potentially significant backreactions on the oscillating  homogeneous inflaton. Hence, if we scan through the parameter $\alpha$, we may be able to find a point where $\tilde{\phi}$ loses its oscillatory nature when the gradient term kicks in, indicating the onset of backreaction from growing perturbations. In Ref. \cite{Lozanov:2017hjm}, it was found that backreaction is significant in $\alpha$-attractor models for $M\ll M_\text{pl}$, which in the present case corresponds to  $\alpha \ll 1$.\\
\begin{figure}[t]
    \centering
    \includegraphics[scale=0.5]{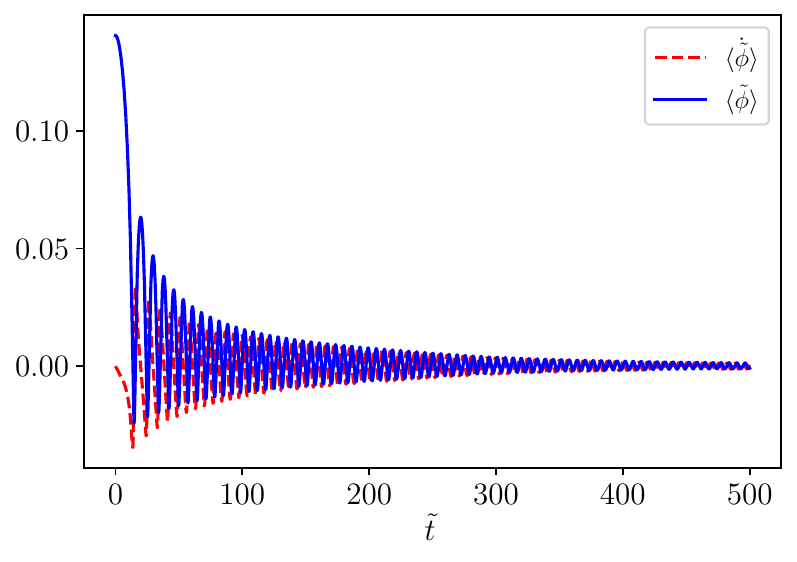}
    \includegraphics[scale=0.5]{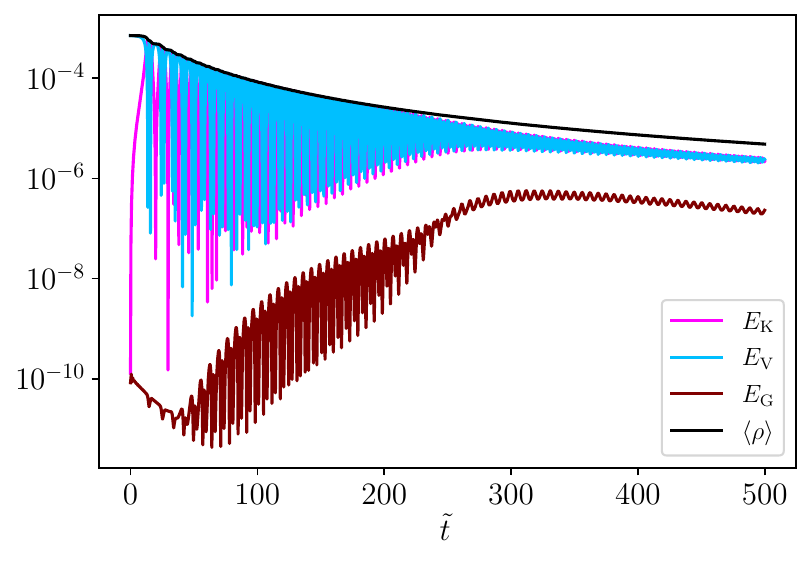}
    \includegraphics[scale=0.5]{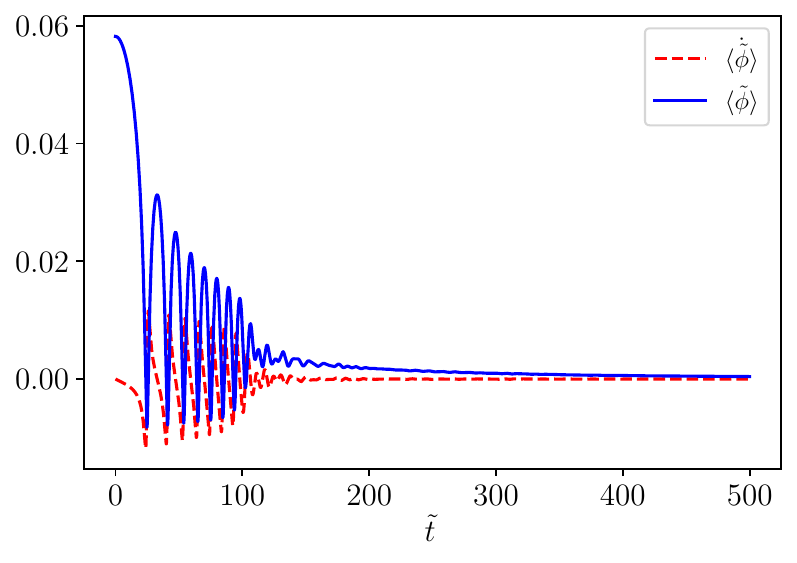}
    \includegraphics[scale=0.5]{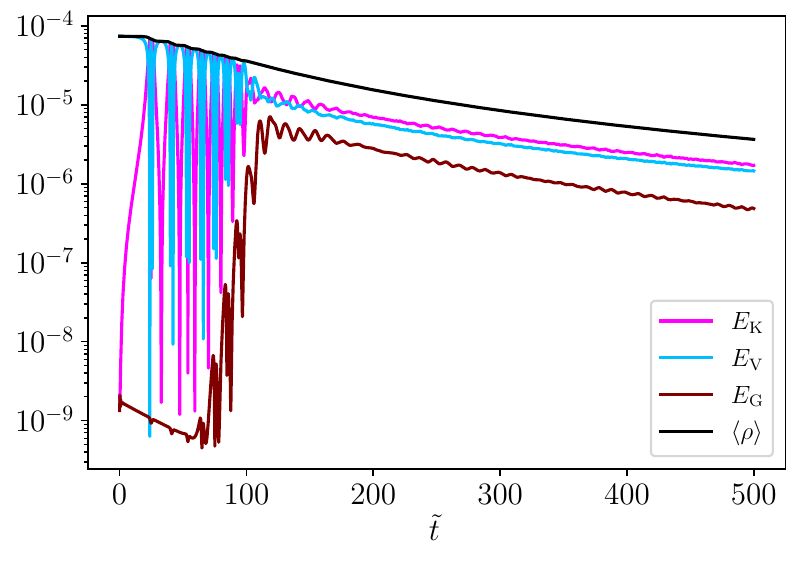}
    \includegraphics[scale=0.5]{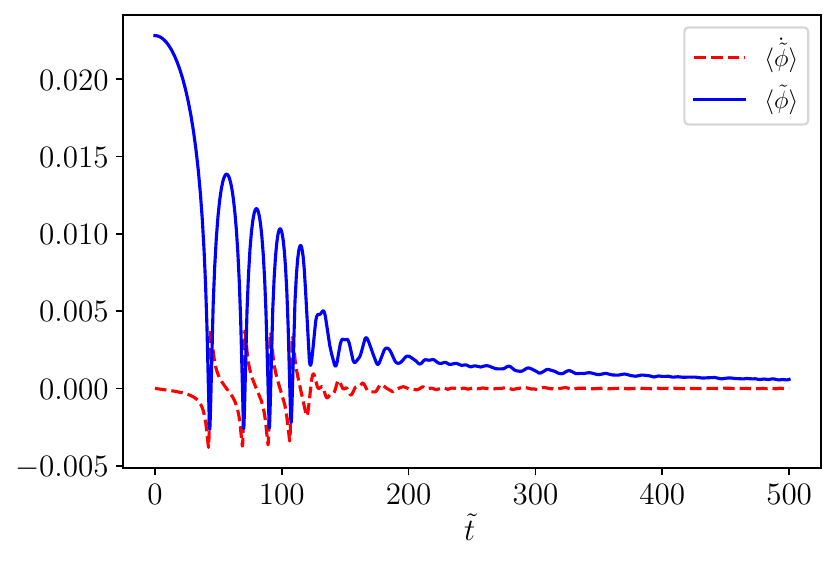}
    \includegraphics[scale=0.5]{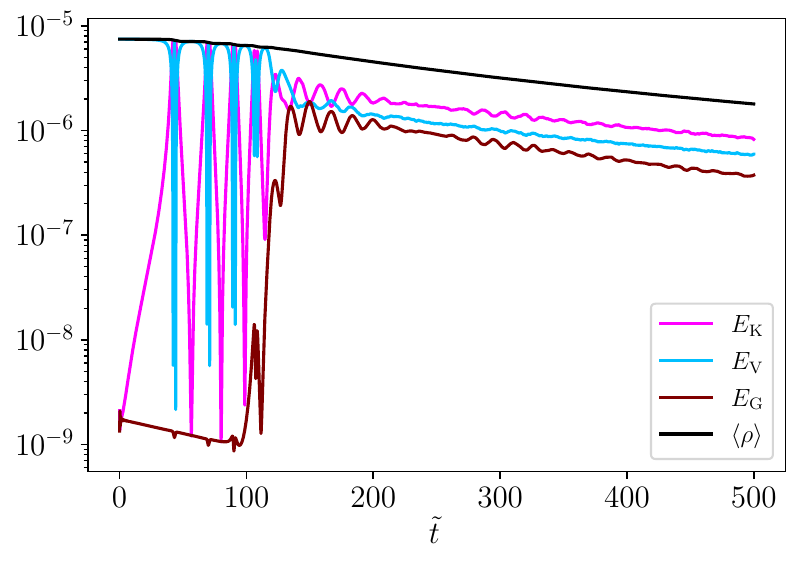}
    \caption{The evolution of volume averaged fields and energies for $\alpha=10^{-3}$ (top), $10^{-4}$ (middle) and $10^{-5}$ (bottom) respectively. The plots show how, with decrease in $\alpha$, the $\tilde{\phi}$ field goes from being oscillatory to a loss of oscillatory nature within $t=100m^{-1}$, suggesting efficient backreaction.}
    \label{fig:fields_and_energies}
\end{figure}
\indent In Fig. (\ref{fig:fields_and_energies}), the evolution of the spatially-averaged field and energy components are shown for parameters $\alpha=10^{-3}$, $10^{-4}$ and $10^{-5}$ in descending order. Since the amount of backreaction is dependent on the strength of the field fluctuations, one tell-tale way of identifying it would to look for the point in time where the gradient energy becomes comparable to the kinetic and potential energies. In other words, $(\widetilde{\grad}\tilde{\phi})^2\sim \dot{\tilde{\phi}}^2, \, \widetilde{V}(\tilde{\phi})$. From the top panel in Fig. (\ref{fig:fields_and_energies}), we see that $E_\text{\tiny G}$ always provides a negligible contribution to the total energy. As a result, there is no noticeable backreaction and the field keeps oscillating with decreasing amplitude. On the other hand, in the middle and bottom panels, we see $E_\text{\tiny G}$ approaching $\mathcal{O}\left( E_{\text{\tiny K}} \right)$ and $\mathcal{O}\left( E_{\text{\tiny V}} \right)$ which, consequently, appears as a backreaction on the evolution of $\tilde{\phi}$. This can be seen in the loss of the oscillatory nature of the field at around $t=100m^{-1}$, from which point $\phi$ maintains a steady decay. Although it may not be feasible to  exactly pin down  the value of $\alpha$ for which backreaction can not be neglected, we find that by narrowing down the range between $5\times 10^{-4}$ and $10^{-4}$, it is possible to infer when the backreaction starts becoming prominent. For this, we refer to Fig. (\ref{fig:backreaction_start}) where the spatially-averaged fields for $\alpha=2.5\times 10^{-4}$ and $2\times 10^{-4}$ are plotted. Although we observe that oscillatory nature of both parameters get suppressed beyond $t=100m^{-1}$, it is for $\alpha<2\times 10^{-4}$ that this effect starts becoming prominent. Hence, the inflaton evolution can be described as a progression of the following stages (for relevant values of $\alpha$)
\begin{enumerate}
    \item Oscillations -- During this period the inflaton maintains its coherent oscillatory nature, and hence the inhomogeneities,   being negligible,  have not started affecting the dynamics.
    \item Transition -- This is a relatively brief period during which the enhanced field fluctuations begin to backreact on the oscillating inflaton condensate, although not strong enough to completely quench the oscillations. It is during this period that the gradient energy $E_\text{\tiny G}$ starts growing appreciably towards $\mathcal{O}(E_\text{\tiny K})$ and $\mathcal{O}(E_\text{\tiny V})$.
    \item Backreaction -- This is the stage where the enhanced field fluctuations backreact onto the background evolution efficiently, thereby significantly changing its dynamics. As such, the inflaton evolution is no longer oscillatory. Moreover, The backreaction also shuts off the enhancement of the field fluctuations, eventually leading to a truncation in the growth of $E_\text{\tiny G}$.
\end{enumerate}

\subsection{Power spectra and nonlinear structure formation}\label{sec:pspectra_and_structure_formation}
\begin{figure}[t]
    \centering
    \includegraphics[scale=0.75]{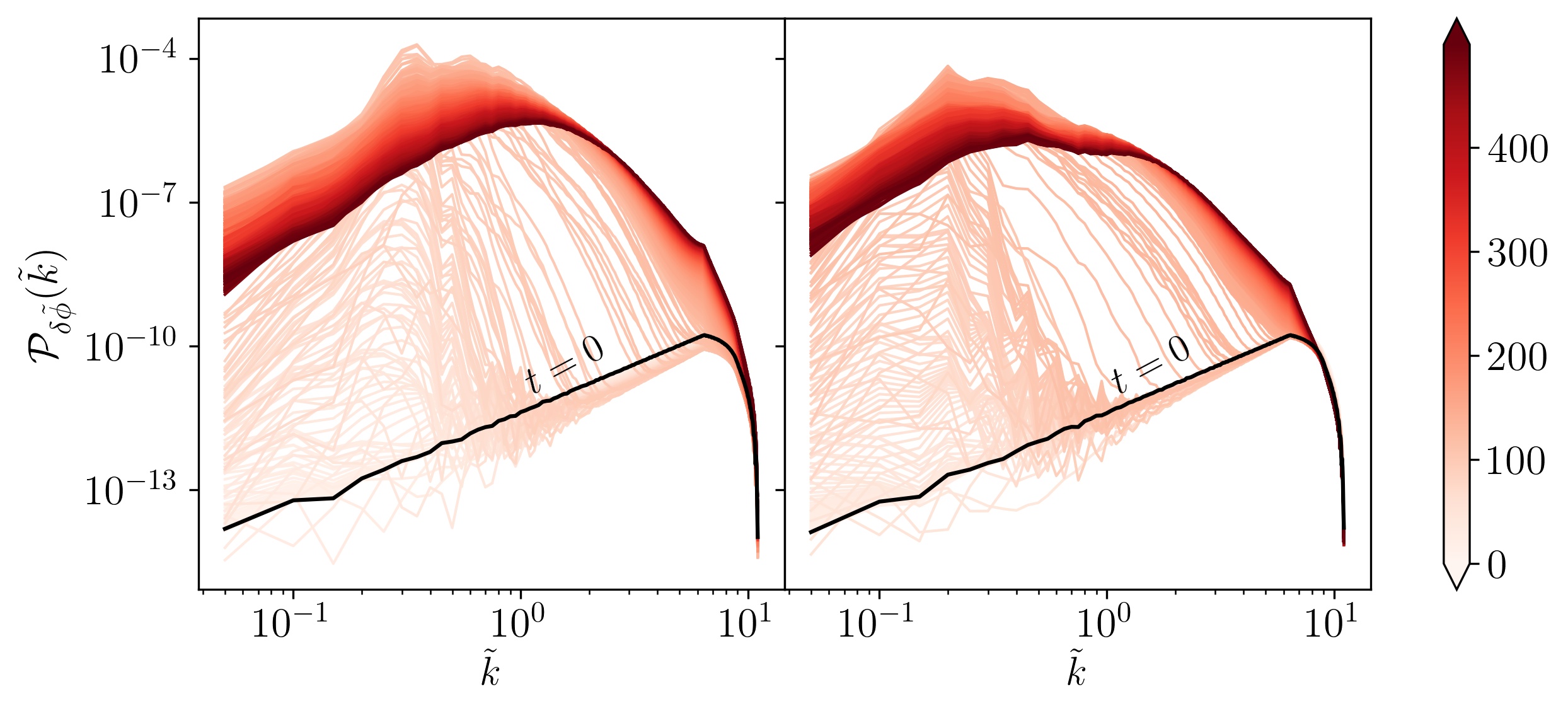}
    \caption{Power spectra of field fluctuations for $\alpha=10^{-4}$ (left panel) and $\alpha=10^{-5}$ (right panel) are shown in the figure with the colorbar representing the progression of time. The solid black lines correspond to the initial power spectra. In these plots $\tilde{k}=k/am$.}
    \label{fig:power_spectra}
\end{figure}
\begin{figure}
    \centering
        \includegraphics[scale=0.5]{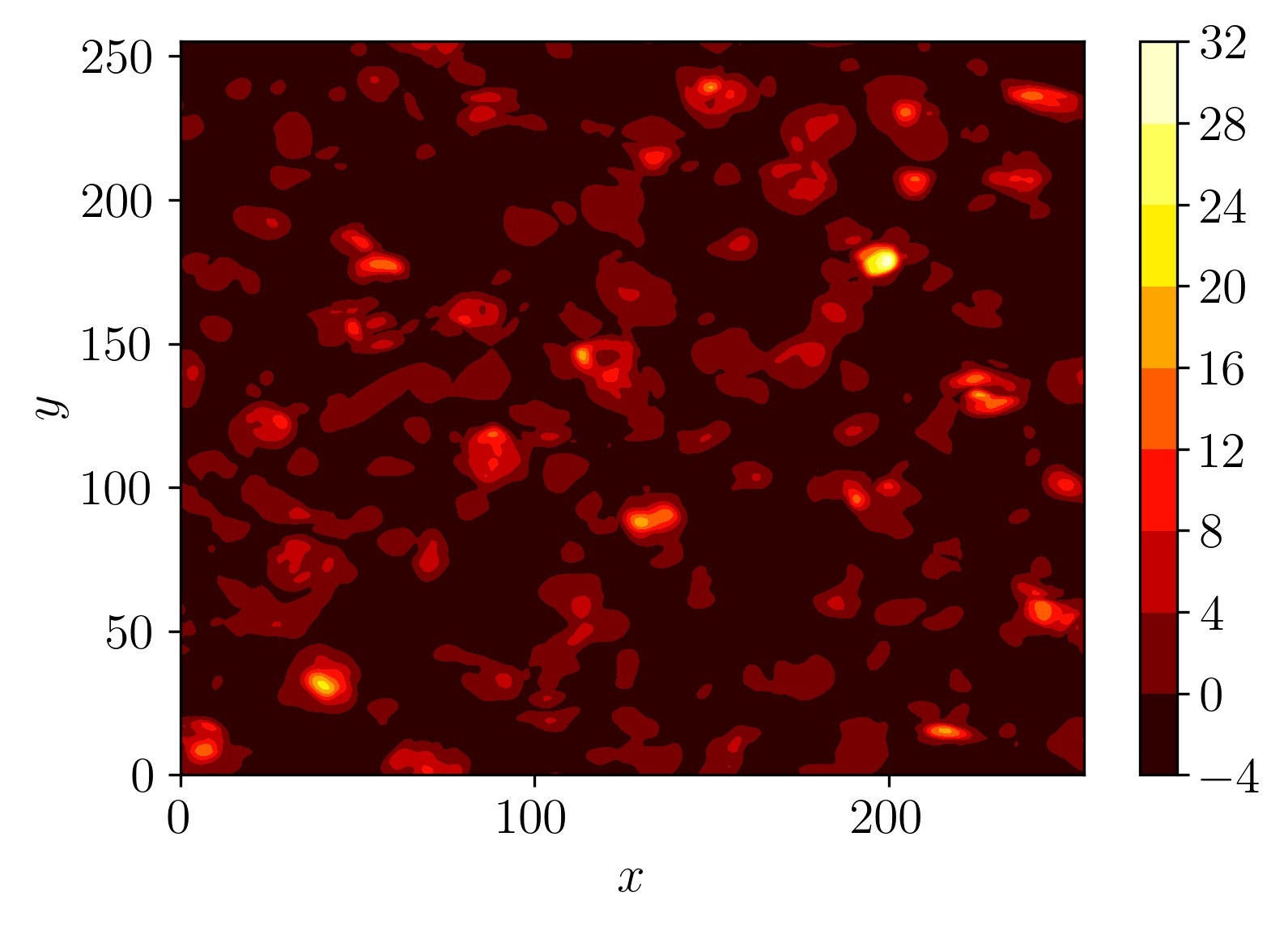}
        \includegraphics[scale=0.5]{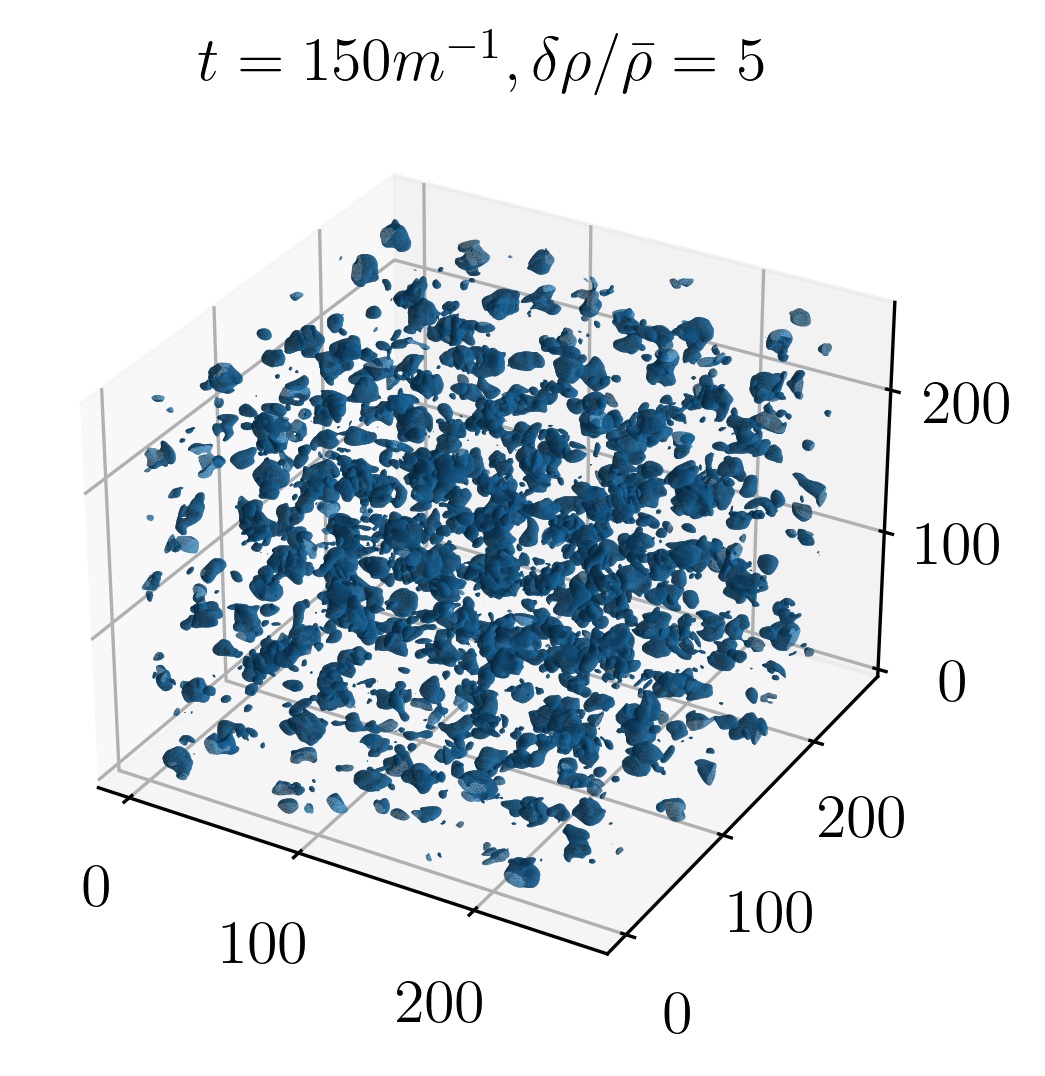}
        \includegraphics[scale=0.5]{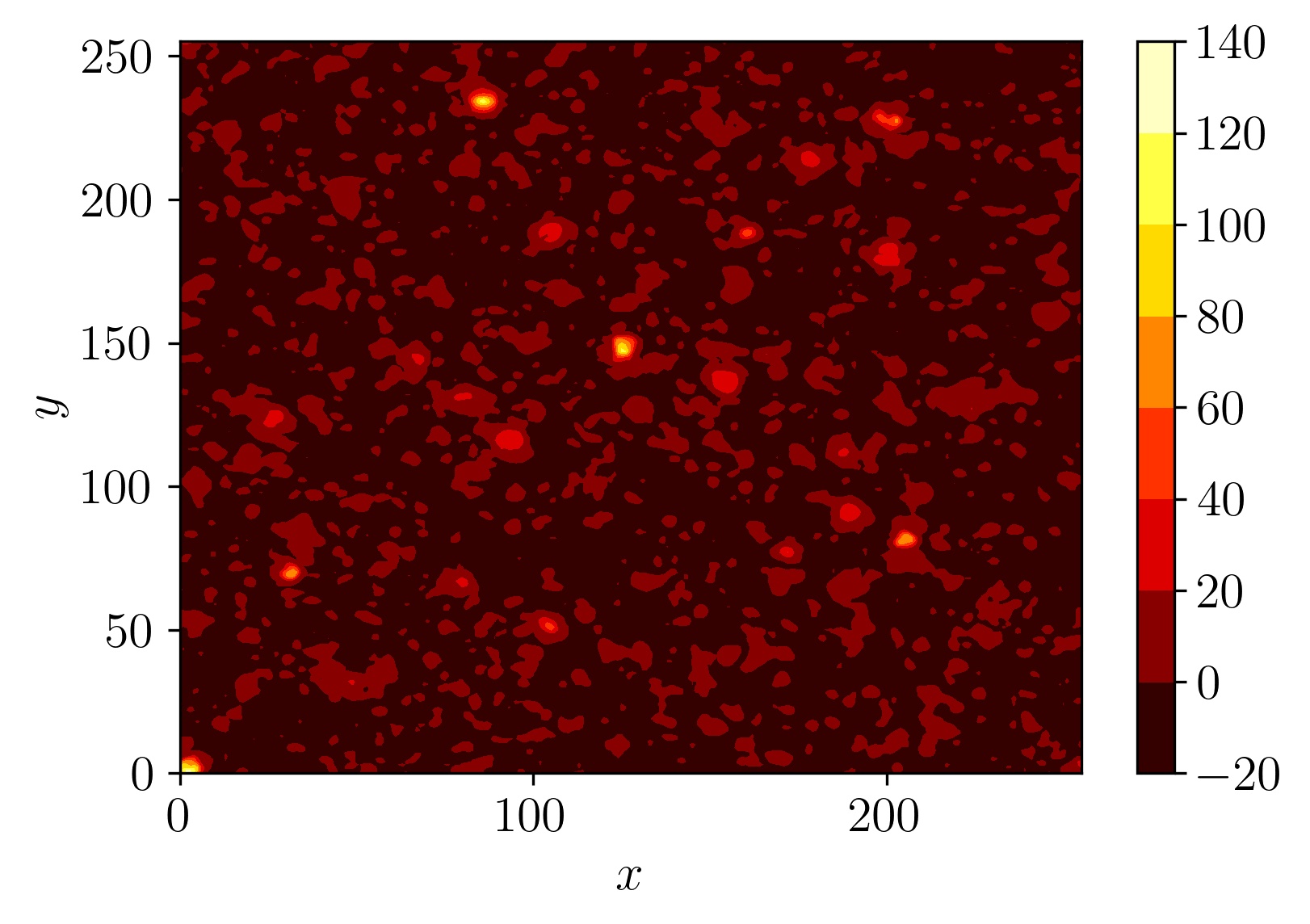}
        \includegraphics[scale=0.5]{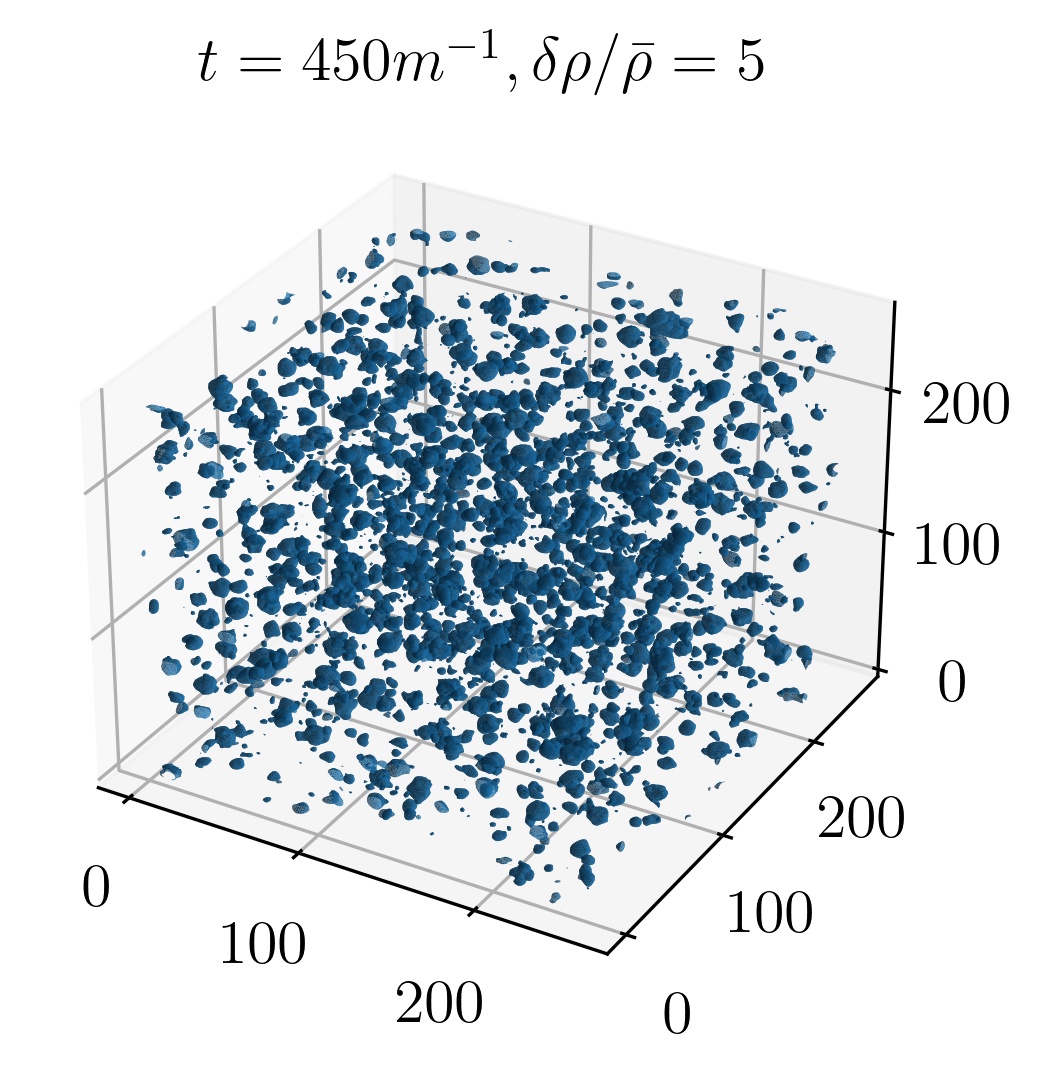}
        \includegraphics[scale=0.5]{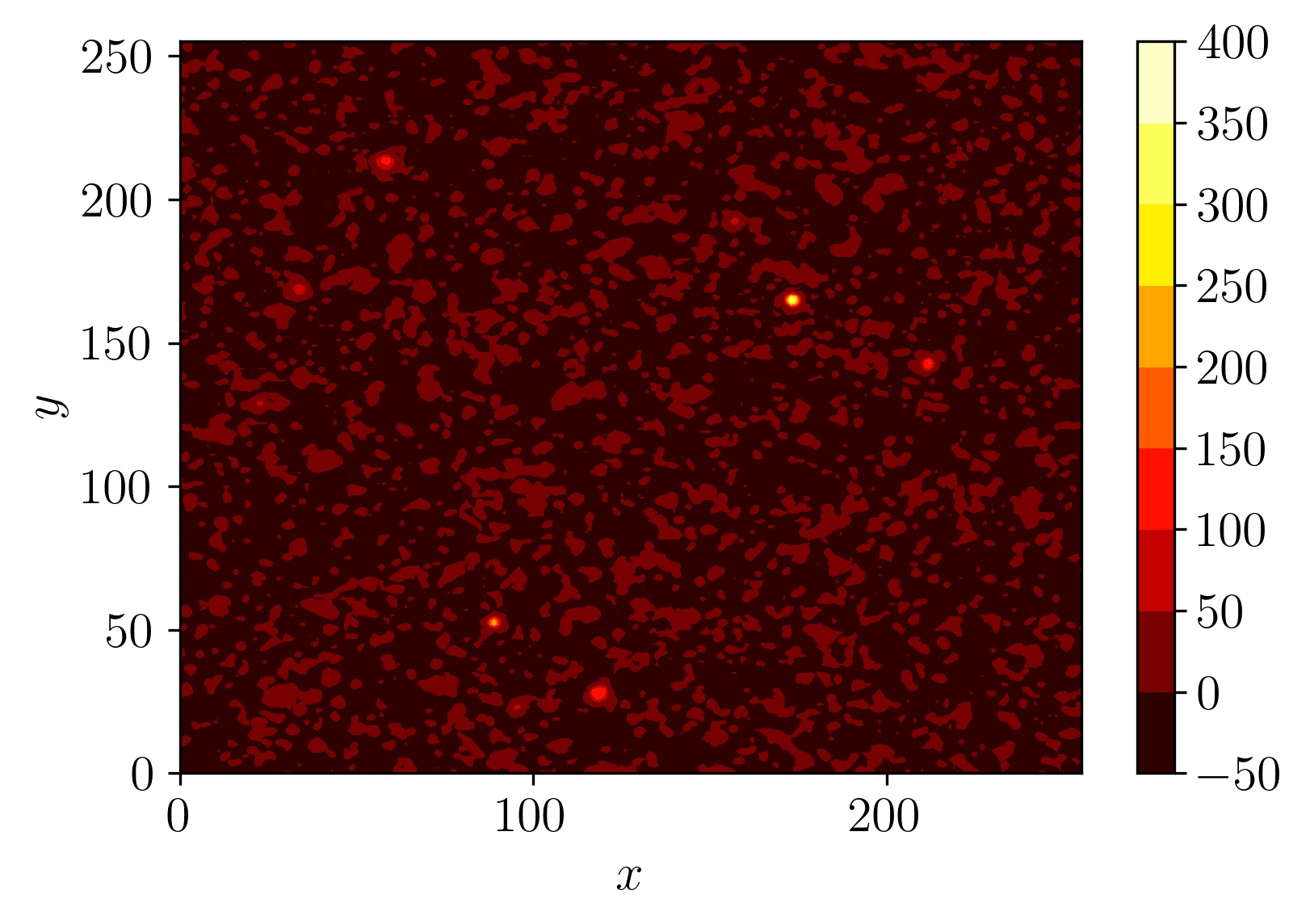}
        \includegraphics[scale=0.5]{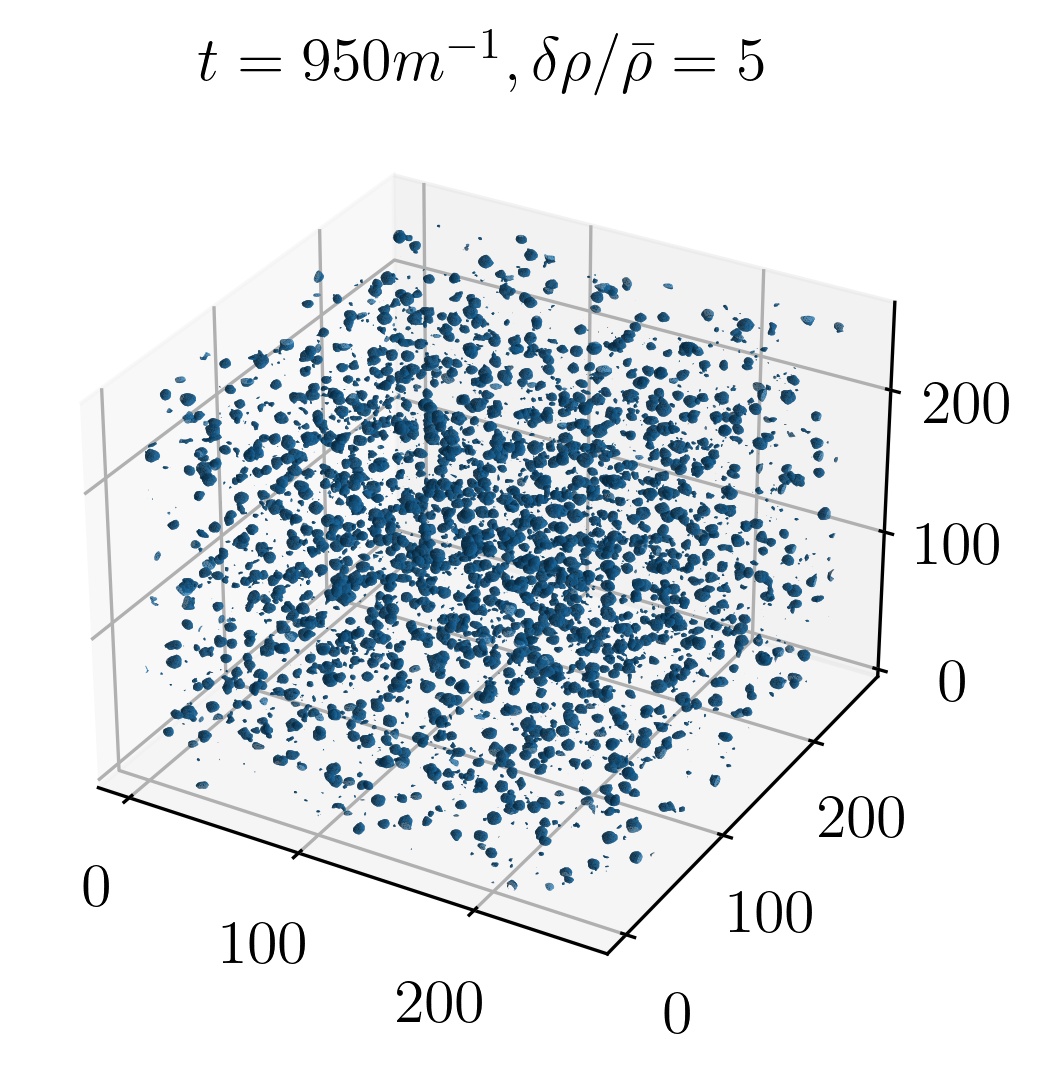}
    \caption{The density contrast $\frac{\delta\rho}{\bar{\rho}}$ across a $2d$ slice (left panel) and the $3d$ energy isosurfaces for $\frac{\delta\rho}{\bar{\rho}}=5$ (right panel) are shown here for $\alpha=10^{-4}$ with $N=256^3$ grid points. The top, middle and bottom panels represent $t=150m^{-1}$, $450m^{-1}$ and $950m^{-1}$ respectively. The colorbars next to the $2d$ slices represent the density contrast magnitude and boxes have physical size $L=125.6m^{-1}$. The sides of these grids and boxes are in comoving lengths with the spatially shrinking overdensities representing oscillons of fixed physical size.}
    \label{fig:3dplots_1}
\end{figure}
\begin{figure}[t]
    \centering
        \includegraphics[scale=0.5]{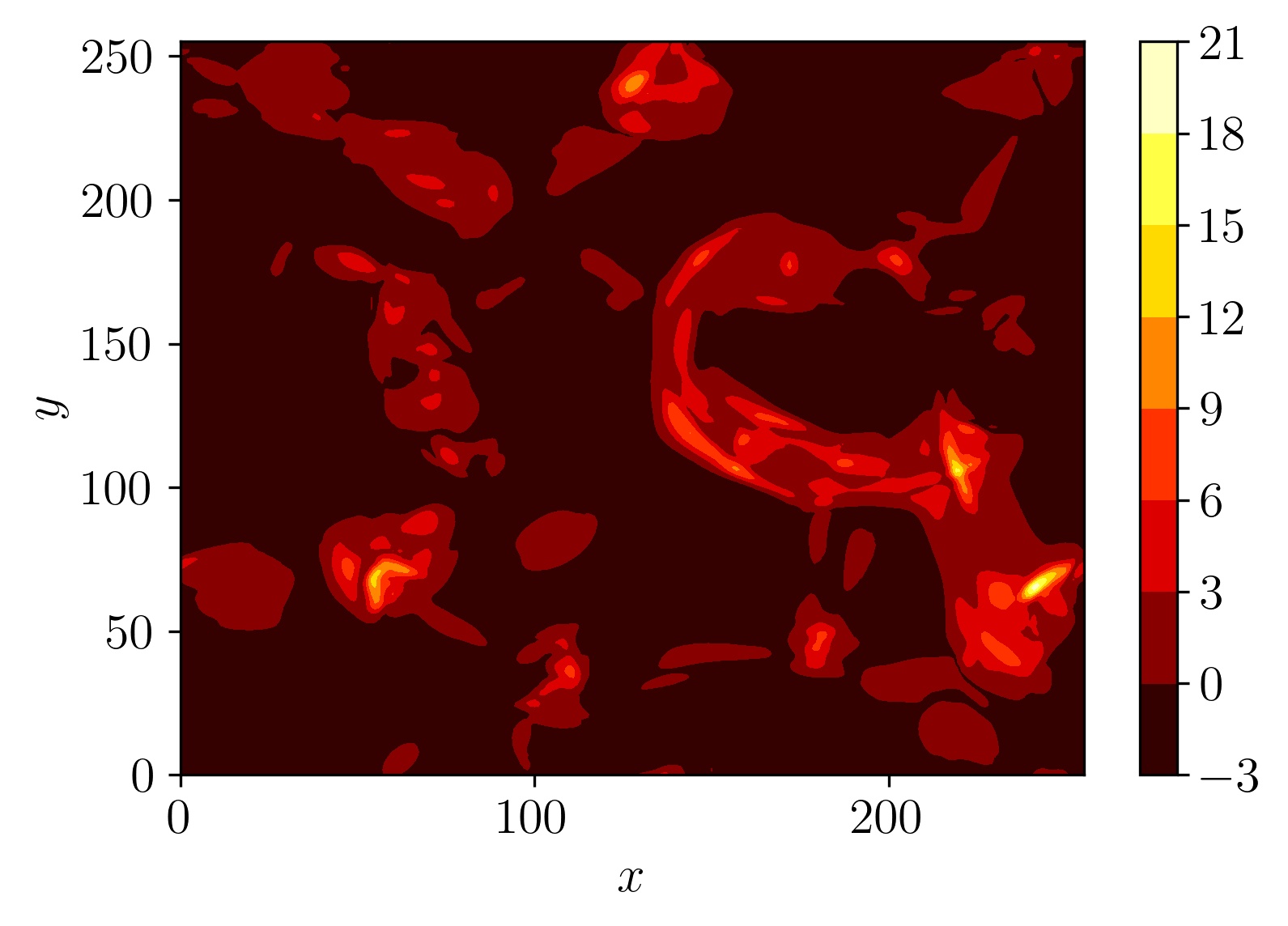}
        \includegraphics[scale=0.5]{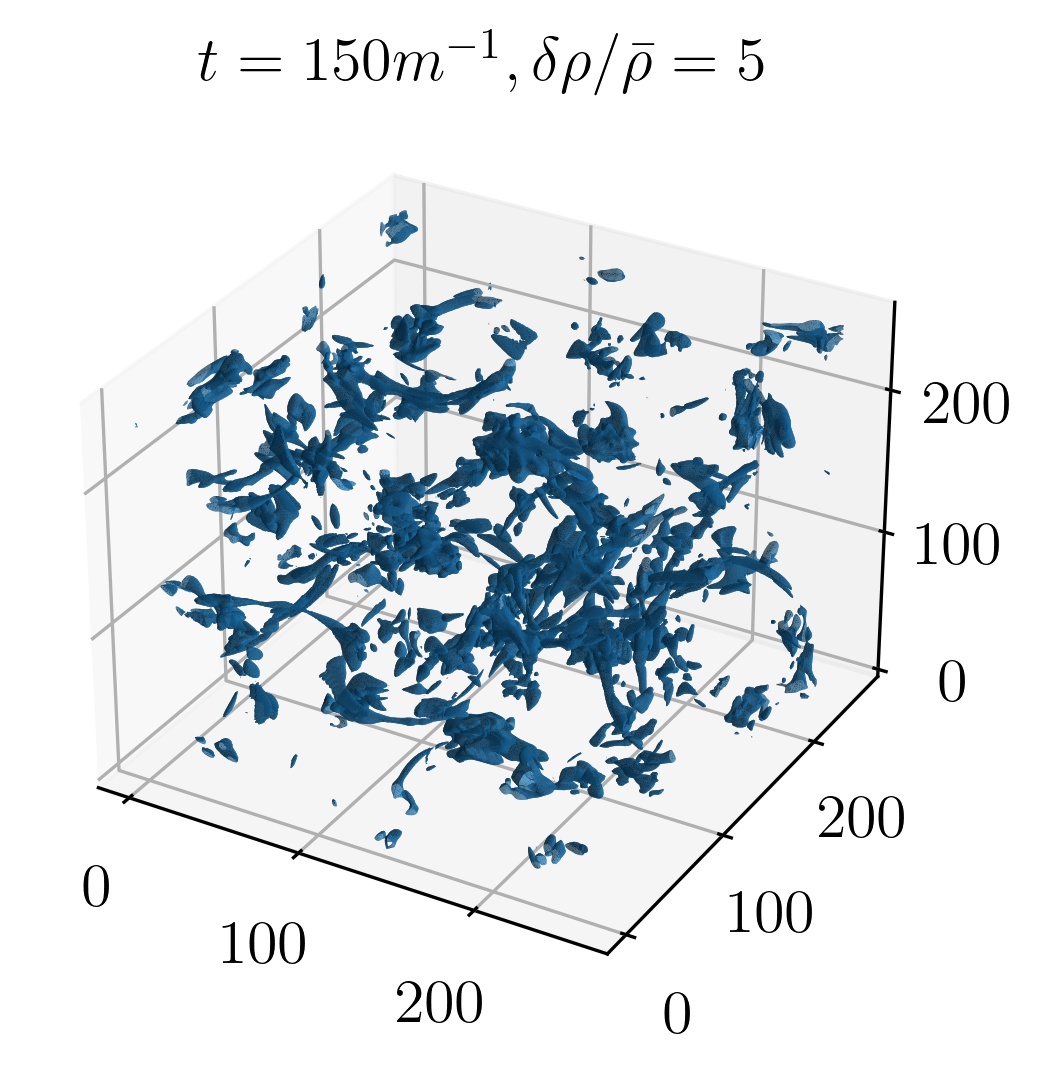}
        \includegraphics[scale=0.5]{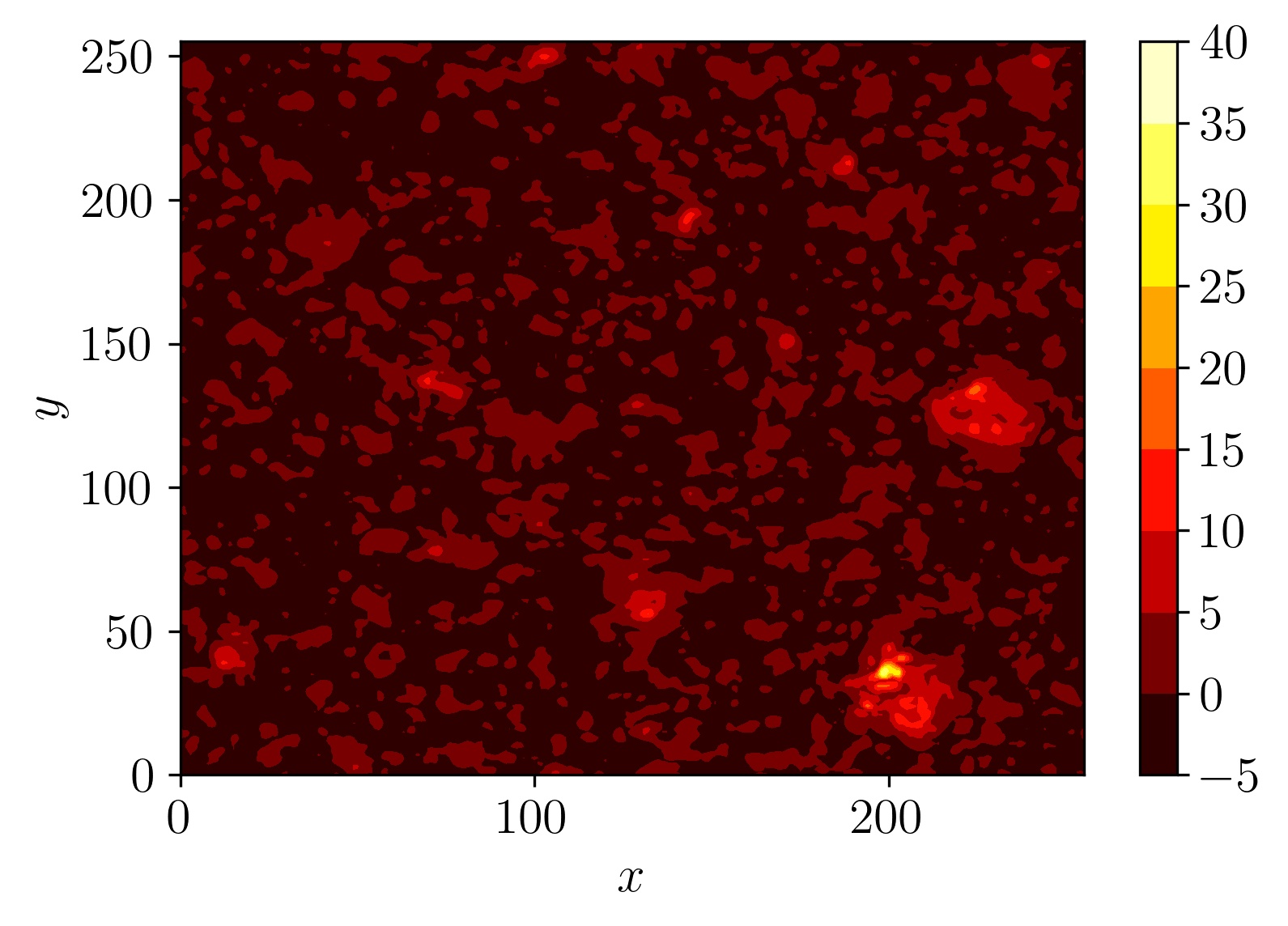}
        \includegraphics[scale=0.5]{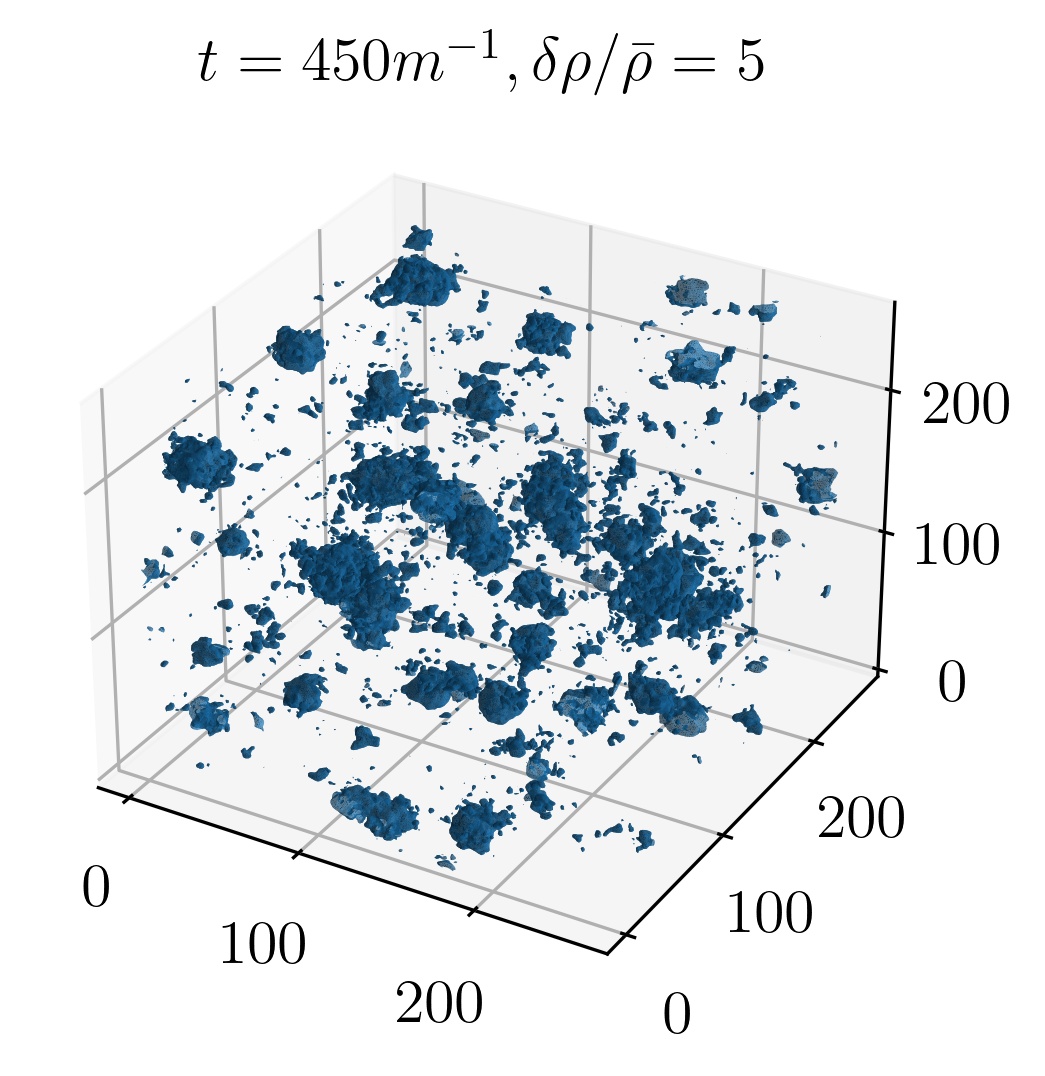}
        \includegraphics[scale=0.5]{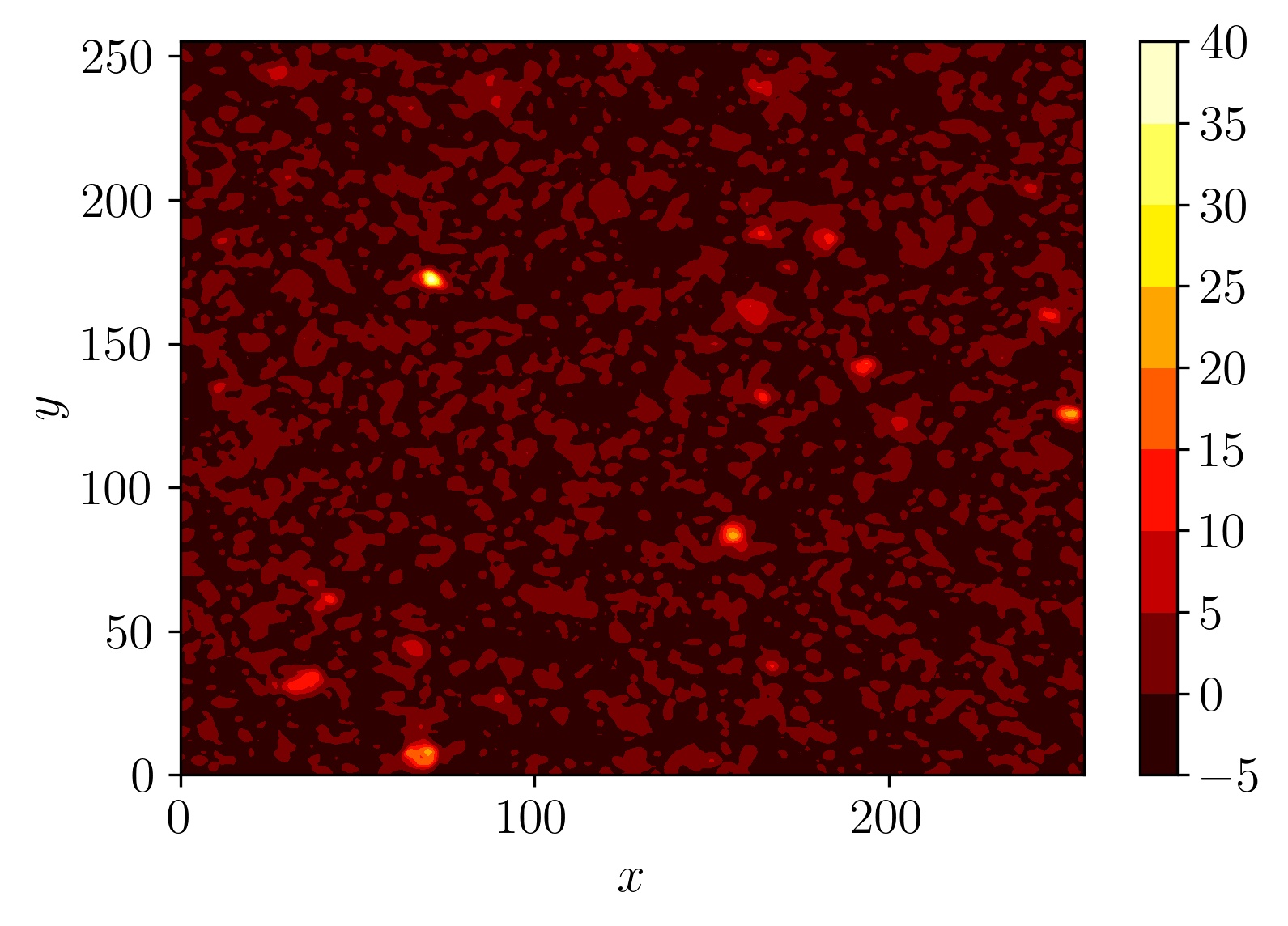}
        \includegraphics[scale=0.5]{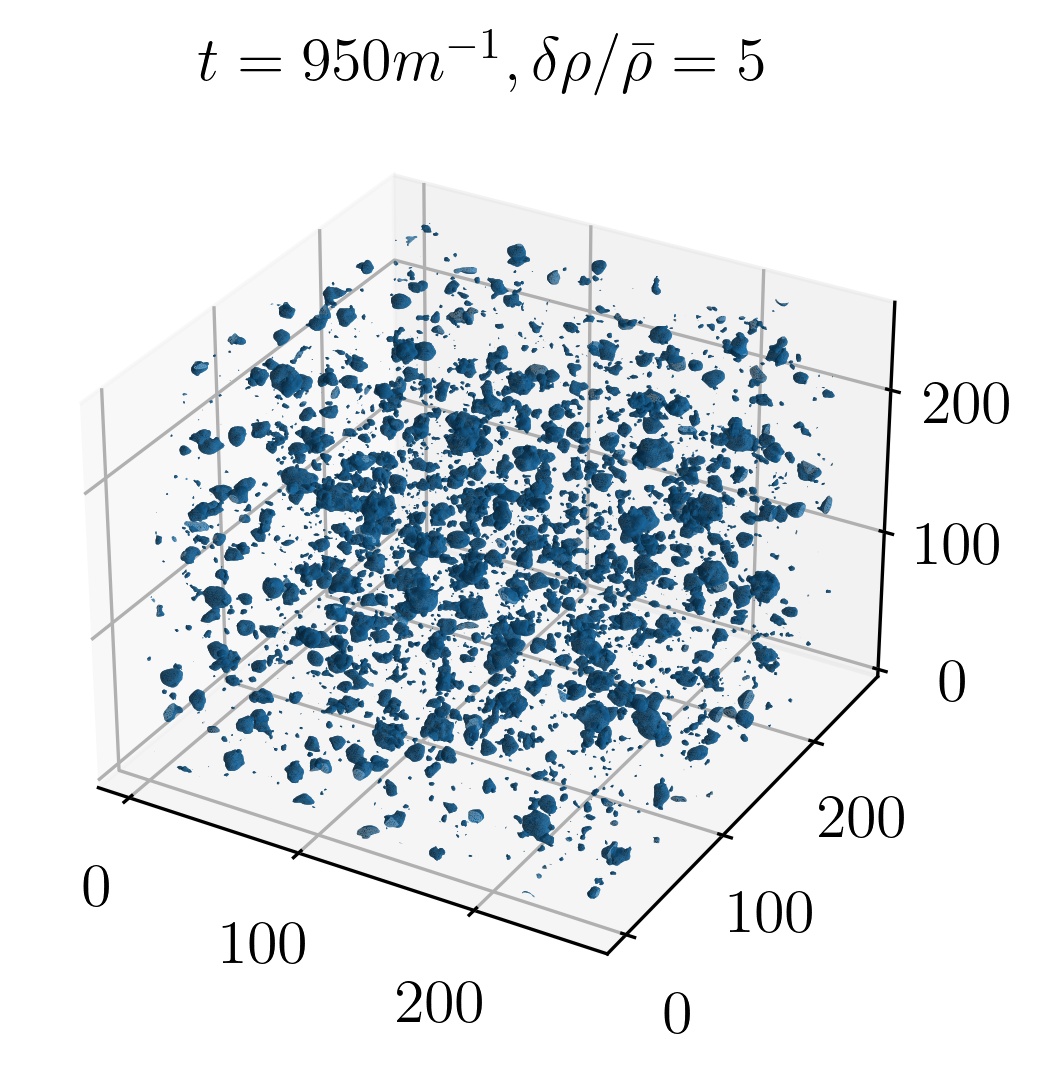}
    \caption{The density contrast $\frac{\delta\rho}{\bar{\rho}}$ across a $2d$ slice (left panel) and the $3d$ energy isosurfaces for $\frac{\delta\rho}{\bar{\rho}}=5$ (right panel) are shown here for $\alpha=10^{-5}$ with $N=256^3$ grid points. The top, middle and bottom panels represent $t=150m^{-1}$, $450m^{-1}$ and $950m^{-1}$ respectively. The colorbars next to the $2d$ slices represent the density contrast magnitude and boxes have physical size $L=125.6m^{-1}$. The sides of these grids and boxes are in comoving lengths with the spatially shrinking overdensities representing oscillons of fixed physical size.}
    \label{fig:3dplots_2}
\end{figure}

We now consider the growth of perturbations, comparing the lattice data for the power spectrum of inflaton fluctuations $\mathcal{P}_{\delta\phi}(k)$ with our expectations from the linear analysis presented in Sec. \ref{sec:Floquet}. In Fig. (\ref{fig:power_spectra}), the power spectra for the field fluctuations are shown for parameters $\alpha=10^{-4}$ (left panel) and $10^{-5}$ (right panel) with the colorbar indicating the passage of time from $t=0$ to $500m^{-1}$. The solid black lines represent the power spectrum at $t=0$. In the figures we see that the smaller modes $\left( k\lesssim 0.5m \right)$ are the first to get excited since, according to the Floquet charts, these correspond to the physical $k$-modes which enter the broad resonance band first. Subsequently, the peaks become broader, being shifted towards larger $k$-modes since these are the ones that enter the broad resonance regime at later times -- after the low $k$-modes have already passed through. \\
\indent Next we study the spatial configurations of inhomogeneities. \textsf{$\mathcal{C}\text{osmo}\mathcal{L}\text{attice}$} is able to generate $3d$ data for the energies $E_\text{\tiny K}$, $E_\text{\tiny V}$ and $E_\text{\tiny G}$ in \textsf{HDF5} files which can be used to calculate the spatial distribution of nonlinear structures by computing the density contrast $\frac{\delta\rho}{\bar{\rho}}$.\footnote{The matter of what level of nonlinearity one should look for is rather arbitrary and different authors have used different thresholds in their works (for example, in Ref. \cite{Hiramatsu:2020obh}, the authors chose this threshold to be $\delta\rho=\bar{\rho}$). In this work, we will specifically look for regions in the lattice where these fluctuations are 5 times the mean energy-density.} In particular, we refer to Figs. (\ref{fig:3dplots_1}) and (\ref{fig:3dplots_2}) for $\alpha=10^{-4}$ and $10^{-5}$ respectively. The plots are arranged as follows
\begin{itemize}
    \item The three rows of the figures are density contrast snapshots at three different times -- namely $t=150m^{-1}$, $450m^{-1}$ and $950m^{-1}$. 
    \item In each row, there are two columns. The left column shows a $2d$ contour plot for $\frac{\delta\rho}{\bar{\rho}}$ at a fixed time with an appropriate colorbar. On the other hand, the right column shows a $3d$ isosurface plot for regions where the overdensities exceed 5 times the mean energy density, \textit{i.e.}, $\delta\rho=5\bar{\rho}$.
\end{itemize}
We observe, through the $2d$ and $3d$ plots of the overdensities, the existence of highly nonlinear structures that form through the instabilities and subsequent growth of the scalar field perturbations. We note that the grid lengths here do not correspond to the actual comoving lengths of the system but the $256$ grid points in each $N_x$, $N_y$ and $N_z$. Moving from $t=150m^{-1}$ to $t=950m^{-1}$, it is seen that these structures shrink. Since the $3d$ lattices are in comoving units, the shrinking is consistent with the formation of overdensities of constant physical size (see Appendix \ref{sec:proper_size}). With the passage of time, most of these these density fluctuations (or \textit{hot spots}) in the $2d$ slices average around 10 times the mean density with a few isolated peaks. Qualitatively, we also observe differences in the nonlinearities in $\alpha=10^{-4}$ and $10^{-5}$ where, in the latter, such regions start out in elongated configurations before fragmenting into separate lumps. However, whether or not these nonlinear structures can be considered as oscillons is a matter that needs to be investigated further. This can be studied by observing the long-term behavior of the EoS and whether or not $w\rightarrow 0$, in which case the nonlinear structures can be interpreted as oscillons. We recall that the spatially-averaged EoS is defined as 
\begin{equation}
    \widetilde{w}=\frac{\widetilde{P}}{\tilde{\rho}}=\frac{\dot{\tilde{\phi}}^2/2-(\widetilde{\grad}\tilde{\phi})^2/6a^2-\widetilde{V}}{\dot{\tilde{\phi}}^2/2+(\widetilde{\grad}\tilde{\phi})^2/2a^2+\widetilde{V}}
\end{equation}
Furthermore, since the EoS rapidly oscillates compared to the Hubble scale, we also perform a time-average (a moving-average) which we denote by $\langle \widetilde{w}_{\text{\tiny T}} \rangle$. Using the virial theorem, we can show that the time and spatially-averaged EoS behaves as follows \cite{Lozanov:2017hjm}
\begin{equation}\label{eq:EoS}
    \langle \widetilde{w} \rangle_{\text{\tiny T}}=\frac{1}{3}+\frac{2}{3}\frac{\left( n-2 \right)}{\left( n+1 \right)+\frac{\langle (\tilde{\grad}\tilde{\phi})^2/a^2 \rangle_{\text{\tiny T}}}{\langle \tilde{V} \rangle_{\text{\tiny T}}}}
\end{equation}
\begin{figure}
    \centering
    \includegraphics[scale=0.5]{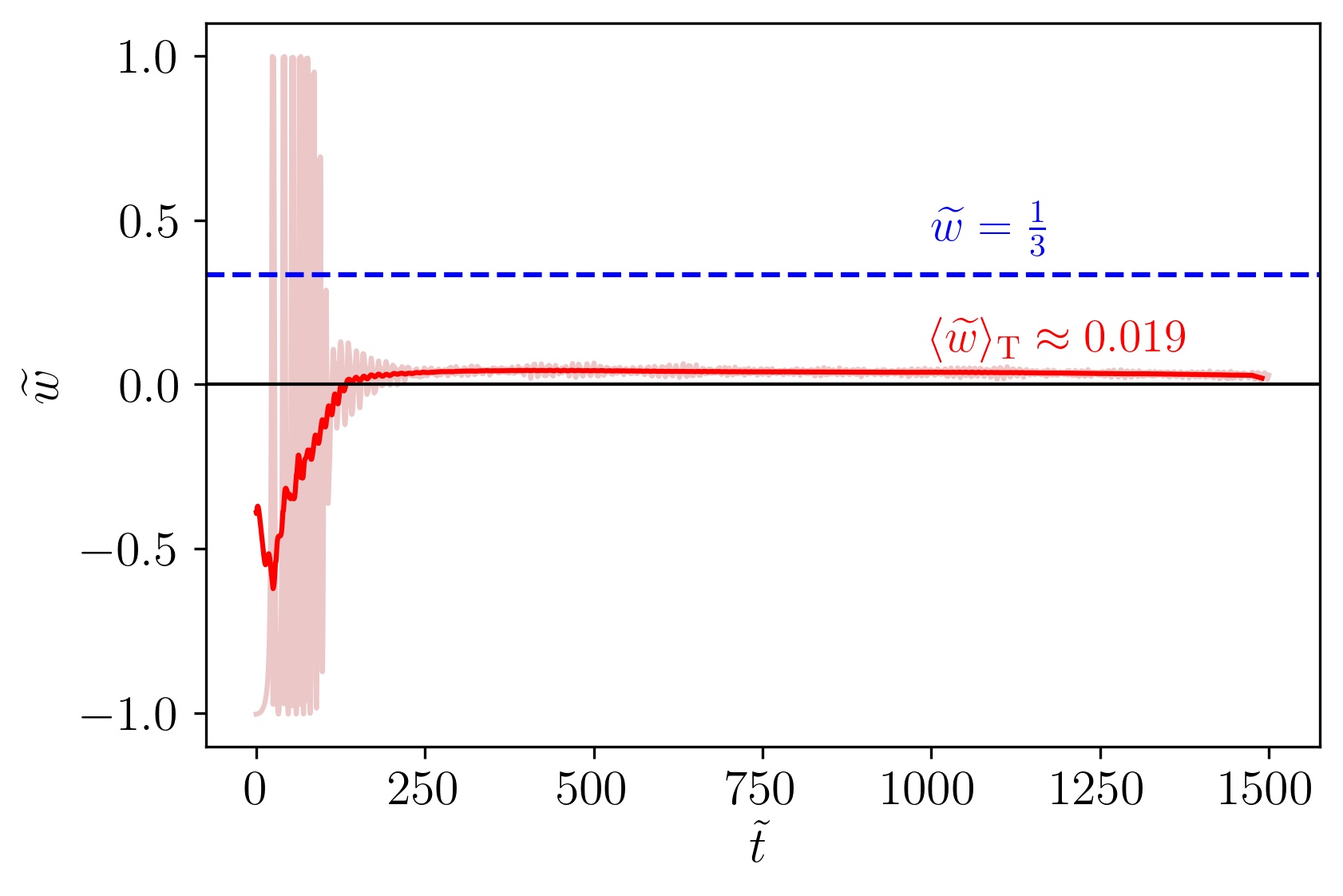}
    \includegraphics[scale=0.5]{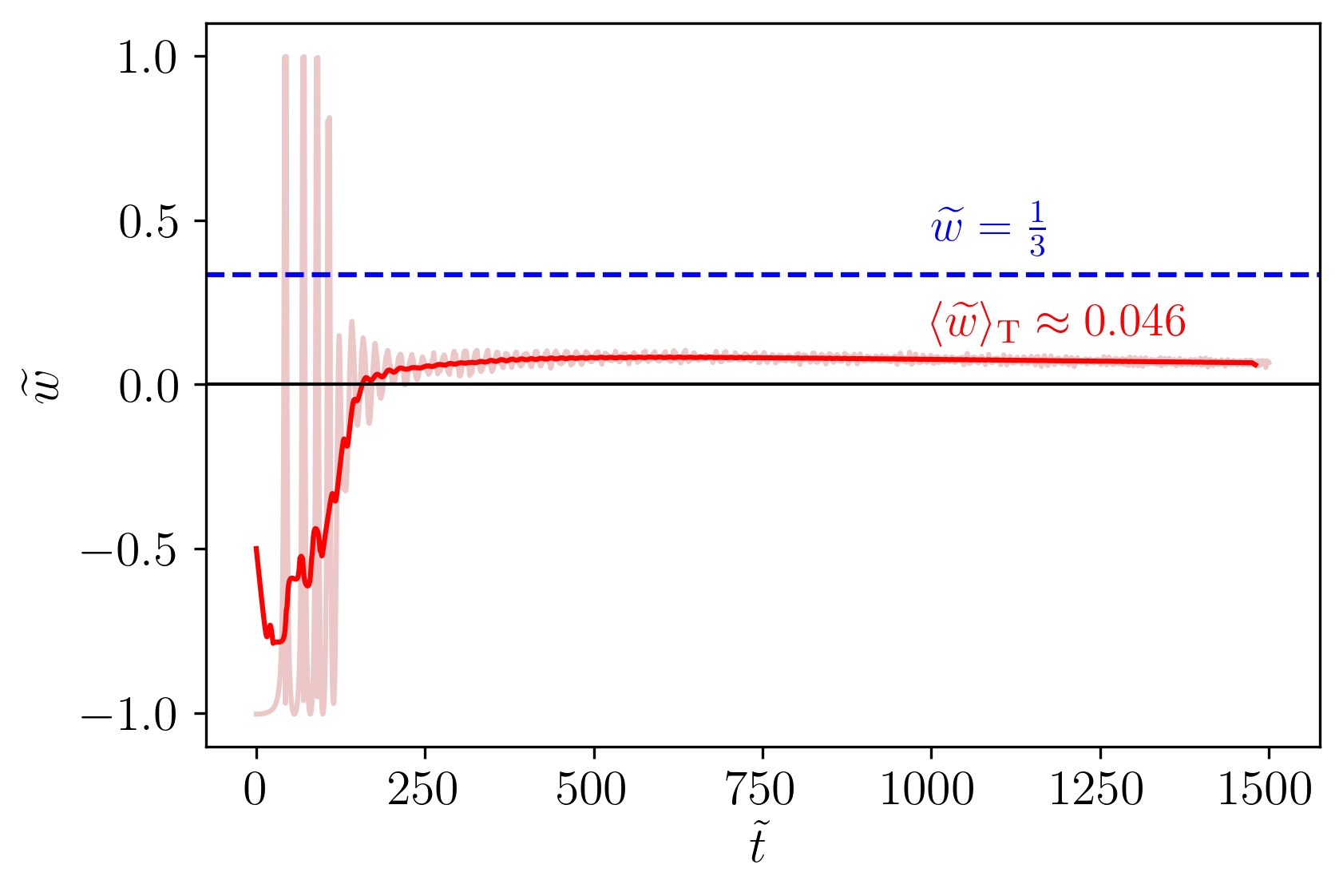}
    \caption{The plots show the evolution of the spatially-averaged EoS $\widetilde{w}$ for $\alpha=10^{-4}$ (left panel) and $\alpha=10^{-5}$ (right panel). The solid red lines are obtained by performing a moving average of $\widetilde{w}$, over a suitably defined window, that we define as a time-average $\langle \widetilde{w} \rangle_{\text{\tiny T}}$. The results of these plots have been generated for $N=128^3$.}
    \label{fig:EoS}
\end{figure}\\
for potentials which are $V\propto \phi^{2n}$ near their minima which, in the absence of inhomogeneities, yields the famous result $ \langle \widetilde{w} \rangle_{\text{\tiny T}} = (n-1)/(n+1)$ (see Ref. \cite{Turner:1983he}). However, numerical simulations indicate that for all $n > 1$, contrary to the homogeneous case,  the average EoS turns out to be $ \langle \widetilde{w} \rangle_{\text{\tiny T}} \to 1/3 $. In any case,  the inflaton fluctuations do not behave like pressureless matter for $n > 1$. If the gradient energy always plays a significantly subdominant role such that $E_\text{\tiny G}\ll E_\text{\tiny V}$, then $\langle \widetilde{w} \rangle_{\text{\tiny T}}\simeq 0$ for $n=1$, indicating a matter-dominated stage. For the $n=1$ case of the E-model, the leading order behavior of the potential is quadratic, hence we expect the average EoS to be close to zero at early times when the backreaction effects are negligible.\\
\indent We expect small deviations from $\langle \widetilde{w} \rangle_{\text{\tiny T}} = 0$ at intermediate times because of a nonvanishing contribution arising from the gradient energy. Observing the plots on the right panel of Fig. (\ref{fig:fields_and_energies}), it is evident that $E_\text{\tiny G}$ is not completely negligible compared to $E_\text{\tiny V}$. As a result, for such intermediate times, we expect the average EoS to be very small, but nonzero nonetheless. However, since $E_\text{\tiny G}$ falls off faster than $E_\text{\tiny V}$ with time, the average EoS is expected to asymptotically approach that of pressureless matter. In Fig. (\ref{fig:EoS}), the average EoS are plotted for the two $\alpha$ parameters. In the plots, the transparent red curves represent only the spatially-averaged EoS $\widetilde{w}$ while the solid red ones include a further time average. It is clear that the nonlinearities that form from preheating behave as pressureless matter which, nevertheless, contains a non-negligible fraction that can constitute of components other than matter. This is evident from the fact that $\langle \widetilde{w} \rangle_{\text{\tiny T}}$, within the provided range, has a small positive value. Moreover, the average EoS for $\alpha=10^{-5}$ is slightly larger in this range and can be understood from how the gradient energy behaves for this parameter. Regardless of this, there is evidence that $\langle \widetilde{w} \rangle_{\text{\tiny T}}$ should approach zero if a longer-term time evolution is considered since the $E_\text{\tiny G}$ curve diverges away from the one for $E_\text{\tiny V}$. These nonlinearities can be regarded as oscillons characterized by a spherically symmetric profile $\phi_{\text{osc}}(t,r)$ peaked at the center and monotonically decaying to zero away from the center. Such profiles should be similar to the secant-type oscillon profiles derived using the small amplitude analysis in Sec. \ref{sec:small_amp}. \\
\indent Finally, we can estimate the fraction of the energy density $f_\text{osc}$ locked in oscillons. A simple prescription for calculating it is given by \cite{Amin:2011hj}

\begin{equation}
    f_\text{osc}\equiv\frac{E_\text{osc}}{E_\text{tot}}=\frac{ \int_{ \delta\rho\gtrsim 2\bar{\rho} }\dd^3\bm{x}\rho(\bm{x},t) }{ \int\dd^3\bm{x}\rho(\bm{x},t) }
\end{equation}
where, in the numerator, the $3d$ energy distribution is integrated for points where the density contrast exceeds two. Since the $3d$ data are over a $256^3$ grid, a straightforward volume integral can prove to be computationally taxing. However, the integrals can be solved very efficiently using a straightforward Monte Carlo integration implementation. With Monte Carlo integration, the integrand
\begin{equation}
    I=\int\dd^3\bm{x}\rho(\bm{x},t)
\end{equation}
can be converted to
\begin{equation}
    I\approx \mathcal{V}\frac{1}{N}\sum_{i=1}^{N}\rho(\bm{x}_i,t)
\end{equation}
\begin{figure}
    \centering
    \includegraphics[scale=0.75]{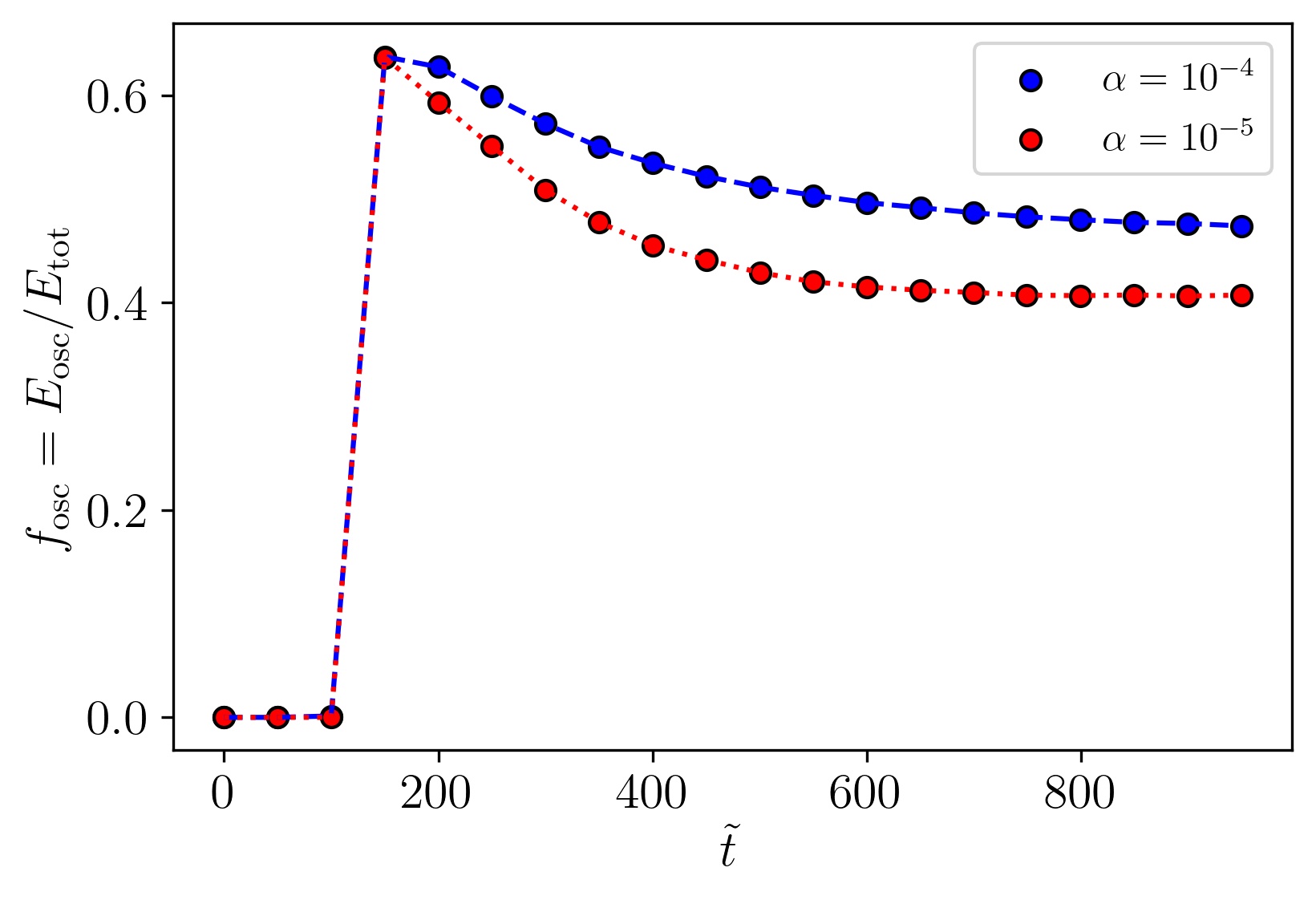}
    \caption{The fraction of energy contained in oscillons $f_\text{osc}$ is shown as a function of time. The fraction attains a non-negligible value from the onset of backreaction, then decays to a constant value which it maintains.}
    \label{fig:f_osc}
\end{figure}
where $\mathcal{V}$ is the volume of the lattice and the sum is over a uniform sampling of points in the lattice. Evaluating the integrals for the two parameters, we summarize the results for $f_\text{osc}$ in Fig. (\ref{fig:f_osc}). They indeed demonstrate that there is very little nonlinear structure formation before the onset of backreaction, at which point $f_\text{osc}$ abruptly shoots up to $60\%$. After this initial rise, the fraction decays to a constant value such that $f_\text{osc}\gtrsim 40\%$ for both parameters. The difference between the values of $f_\text{osc}$ can be explained by the choice of the threshold for the density contrast since the way the overdensities cluster may be different in the two parameters. Moreover, we can study the $2d$ overdensity plots for another possible answer. In Fig. (\ref{fig:3dplots_1}) and (\ref{fig:3dplots_2}), we observe that, although the hotspots are similar in distribution, the ones for $\alpha=10^{-4}$ feature more prominent peaks at certain locations compared to $\alpha=10^{-5}$. Although not conclusive why there should be different oscillon fractions arising from these two parameters, these provide two possible explanations upon examination of the results.  Furthermore, it is possible that the  increased abundance of  oscillons observed in Fig. (\ref{fig:f_osc}), as the efficiency of oscillon production  is lowered due to backreaction effects,   might be related to the inadequate UV resolution of our simulations. This is also indicated by the fact that  there is a kink-like
feature at late times (around $\tilde{k}\simeq 6$) in the power-spectra of field fluctuations, as can be seen in Fig.~(\ref{fig:power_spectra}). We are keen on carrying out a detailed  investigation into these issues in a future project.

\section{Discussion}\label{sec:discussion}

We have shown that oscillons can form during the preheating stage of an asymmetric inflationary potential like the $\alpha$-attractor E-model. In particular, we have demonstrated that backreaction of field fluctuations onto the homogeneous field evolution becomes significant for $\alpha\lesssim2\times 10^{-4}$ leading to the  formation of  highly inhomogeneous field configurations composed of  a significant fraction  of long-lived oscillons. This does not imply that oscillons (and  other nonlinearities) cannot form for larger values of $\alpha$ and, in fact, they can. However, the fractional energy density of the oscillons will be sub-dominant and the preheating dynamics will be dominated by the the coherently oscillating inflaton condensate, given the fact that backreaction effects do not kick in for larger values of $\alpha$. In this such a scenario we can leverage long-term gravitational effects. Using the linearly perturbed Einstein equations, we can show that the Bardeen potential is given by
\begin{equation}
    \Psi_{\bm{k}}=\frac{3}{2}\left( \frac{aH}{k} \right)^2\delta_{\bm{k}}
\end{equation}
where $\delta_{\bm{k}}$ is the Fourier transform of the density contrast. Typically for $\delta_{\bm{k}}\ll 1$ and $k\gg aH$, $\Psi_{\bm{k}}\approx 0$ at early times. Nevertheless, one can still expect the formation of nonlinear structure at late times due to the gravitational amplification of metric fluctuations. As briefly mentioned before, the presence of metric perturbations has important consequences for the growth of $\delta\phi_{\bm{k}}$, so much so that even 
for $V(\phi)\sim\phi^2$, for which inflaton self-interaction is absent, Eq. \eqref{eq:phi_pert_metric} takes the form of the Mathieu equation, which is not the case in the absence of metric fluctuations. With this \textit{metric preheating} phenomenon, it has been shown that small scale perturbations are susceptible to parametric resonance which re-enter the horizon during reheating and which are also larger than a characteristic scale given by $\sqrt{3Hm}$ \cite{Jedamzik:2010dq,Easther:2010mr,Martin:2020fgl}. The effect of gravitational clustering on oscillon formation in this parameter space will be  explored in a future work.\\
\indent It is important to note that the lattice simulation results may change upon further refinement of the lattice size. For the majority of this work, the results reflect the use of $N=256^3$, $\tilde{k}_\text{\tiny IR}=0.05$ and $\tilde{k}_\text{\tiny UV}=4$ (ultraviolet cut-off). It is entirely possible that a finer lattice size ($N=512^3$ or $1024^3$) might reveal interesting details about nonlinear structure formation hitherto unseen, although we do not anticipate very large deviations away from the results that have been presented here. As an example, the observation that $f_\text{osc}$ is less for $\alpha=10^{-5}$ may have arisen from a lack of resolution in the $3d$ grids and one may conjecture that refining the grid can reveal finer details and create better parity between the two parameters.\\
\indent Furthermore, one should also question whether such low values of $\alpha$ perform well with CMB constraints arising from Planck and BICEP/Keck. In the large-$N$ limit, and for $\alpha \leq 1$ the $\alpha$-attractors exhibit the following universality class for the CMB observables $n_s$ and $r$ \cite{Kallosh:2013hoa,Kallosh:2013yoa}
\begin{equation}
    n_s\simeq 1-\frac{2}{N},\;\;\;\;\; r\simeq\frac{12\alpha}{N^2}
\end{equation}
where $N=N_\star-N_\text{end}$ denotes the number of $e$-folds of expansion in between the Hubble-exit epoch of the CMB pivot scale $N_\star$ and the end of inflation $N_\text{end}$. For $\alpha\lesssim\mathcal{O}(1)$ and $N=55$, the $\alpha$-attractors provide very robust predictions for such inflationary observables. For example, for $\alpha=1$ (coinciding with the traditional Starobinsky potential) and $N=55$, one finds that $n_s\approx 0.964$ and $r\approx 0.004$, which are in excellent agreement with current CMB constraints \cite{Planck:2018jri,Planck:2018vyg}. However, $\alpha$-attractors can also produce negligibly small levels of tensor fluctuations depending on the smallness of the value of $\alpha$, which can have major implications for the running of the scalar spectral index $\alpha_s$. We recall that the running of the scalar spectral index is defined as
\begin{equation}
    \mathcal{P}_{\zeta}(k)=\underbrace{\mathcal{A}(k_{\star})}_{2.1\times 10^{-9}}\left( \frac{k}{k_{\star}} \right)^{n_s-1+\frac{1}{2}\alpha_s\ln\left( k/k_{\star} \right)+\cdot\cdot\cdot}
\end{equation}
\begin{equation}
    \alpha_s\equiv\frac{\dd n_s}{\dd \ln k}\bigg\lvert_{k=k_{\star}}
\end{equation}
being currently constrained to $\alpha_s=-0.006\pm 0.013$. In Ref. \cite{Easther:2021cpj} it was shown that inflationary models described by the first two slow-roll parameters are now excluded with the latest BICEP3/Keck \cite{BicepKeck:2021ybl,BICEPKeck:2022mhb} bounds on $r$. With the latest data, the analysis found a posterior distribution $\mathcal{P}(N)$ preferring $N\gtrsim 80$ with the two-term slow-roll hierarchy. The situation can be ameliorated by considering the first three slow-roll parameters. However, with this, very low values of $r$  in asymptotically flat potentials result in  a relatively large running of the scalar spectral index, but are still within current observational bounds on $\alpha_s$.
\section{Conclusions}
In this paper, we investigated whether oscillons can form in an asymmetric inflationary potential -- a question which has not been adequately addressed in the existing literature. We used the $\alpha$-attractor E-model as a representative asymmetric potential. Using a $4^\text{th}$-order Taylor expansion of the E-model potential, we analytically {demonstrated} the existence of oscillon-like solutions with a secant-type core. However, the existence of such oscillon-like solutions is not a sufficient condition for oscillon formation and, considering the fact that the 
$4^\text{th}$-order expression only really works well for $\alpha\sim\mathcal{O}(1)$, a full $3d$ lattice treatment is necessary to verify the  formation of nonlinear objects.\\
\indent We performed a detailed lattice study during preheating in the E-model for three different values of $\alpha$ where we showed that effects of backreaction become significant for $\alpha\lesssim 2\times 10^{-4}$. This occurs due to a sharp rise in the field's gradient energy which begins to  modify the evolution of the oscillating inflaton condensate significantly. In the relevant parameter space, localised and highly nonlinear structures were seen to be formed which maintain roughly constant physical sizes as the universe expands in time. Moreover, the average EoS reveal that they are in fact close to being matter-like (with some non-negligible fraction of the energy being locked into radiative modes which tends to vanish in the asymptotic future. With these in mind, we conclude that indeed a significant amount of oscillons  form in this particular example of an asymmetric potential for $\alpha\lesssim 2\times 10^{-4}$. In relation to the estimate given by Eq. \eqref{eq:alpha_estimate} found in Ref. \cite{Kim:2021ipz}, it is seen that the onset of backreaction occurs for much smaller values of $\alpha$, although it does manage to exclude a large portion of the parameter space.\\
\indent We bear in mind, however, that we have restricted ourselves to only studying the scalar field fluctuations and gravitational influences arising from metric perturbations have been ignored. As mentioned in Sec. \ref{sec:discussion}, for values of $\alpha$ where self-resonance is not as efficient, one can look forward to long-term gravitational effects on the system and the eventual formation of nonlinearities in larger proportions than those formed solely from self-resonance. Such a study will be performed in a future work using full numerical relativity \cite{Kou:2019bbc}. Moreover, larger values of $\alpha$ will imply larger values of $r$ which will be more favorable in terms of CMB constraints. This is not to say that a very small value of $r$ poses any serious problems. However, as seen in Ref. \cite{Easther:2021cpj}, very small values of $r$ point towards somewhat larger values of the running of scalar spectral index  $\alpha_s$ which can impinge on current CMB constraints. Additionally, we note that oscillons, though long-lived, are meta-stable and they eventually decay on longer time scales. We intend to carry out a thorough analysis of oscillon decay in our upcoming paper, focusing on possible astrophysical and cosmological implications both in the case of the inflaton field and ultra-light scalar dark matter.

\section{Acknowledgments}
SSM is supported by an STFC Consolidated Grant [Grant No. ST/T000732/1]. We thank Daniel Figueroa for email correspondences regarding the use of \textsf{$\mathcal{C}\text{osmo}\mathcal{L}\text{attice}$} during the initial phase of this project. For the purpose of open access, the authors have applied a CC BY public copyright license to any Author Accepted Manuscript version arising. \\

{\bf Data Availability Statement:} This work is entirely theoretical and has no associated data. The data files for the lattice simulations (with the exception of the $3d$ configuration files) and other codes can be found in the following GitHub repository: \url{https://github.com/RM503/Oscillon_Emodel}.

\appendix 

\section{Deriving $\mathcal{C}\text{osmo}\mathcal{L}\text{attice}$ parameters from inflationary observables}\label{sec:inflationary_parameters}
Here we use CMB constraints on inflationary observables to derive the parameters and initial homogeneous field configurations for $\mathcal{C}\text{osmo}\mathcal{L}\text{attice}$. With an E-model potential of the form
\begin{equation}
    V(\phi)=\frac{3}{4}\alpha m^2 M_\text{pl}^2\left( 1-e^{-\sqrt{\frac{2}{3\alpha}}\frac{\phi}{M_\text{pl}}} \right)^2
\end{equation}
the first slow-roll parameter $\epsilon_\text{\tiny V}$ can be used to determine the field value at the end of inflation, which serves as the initial condition for the simulations. Using the fact that
\begin{equation}
    \epsilon_\text{\tiny{V}}=\frac{M_\text{pl}^2}{2}\left( \frac{\partial_\phi V}{V} \right)^2
\end{equation}
we solve for $\epsilon_\text{\tiny{V}}=1$ to obtain
\begin{equation}\label{eq:phi_end}
    \frac{\phi_\text{end}}{M_\text{pl}}=\sqrt{\frac{3\alpha}{2}}\ln\left( 1+\frac{2}{\sqrt{3\alpha}} \right)
\end{equation}
\begin{figure}
    \centering
    \includegraphics[scale=0.75]{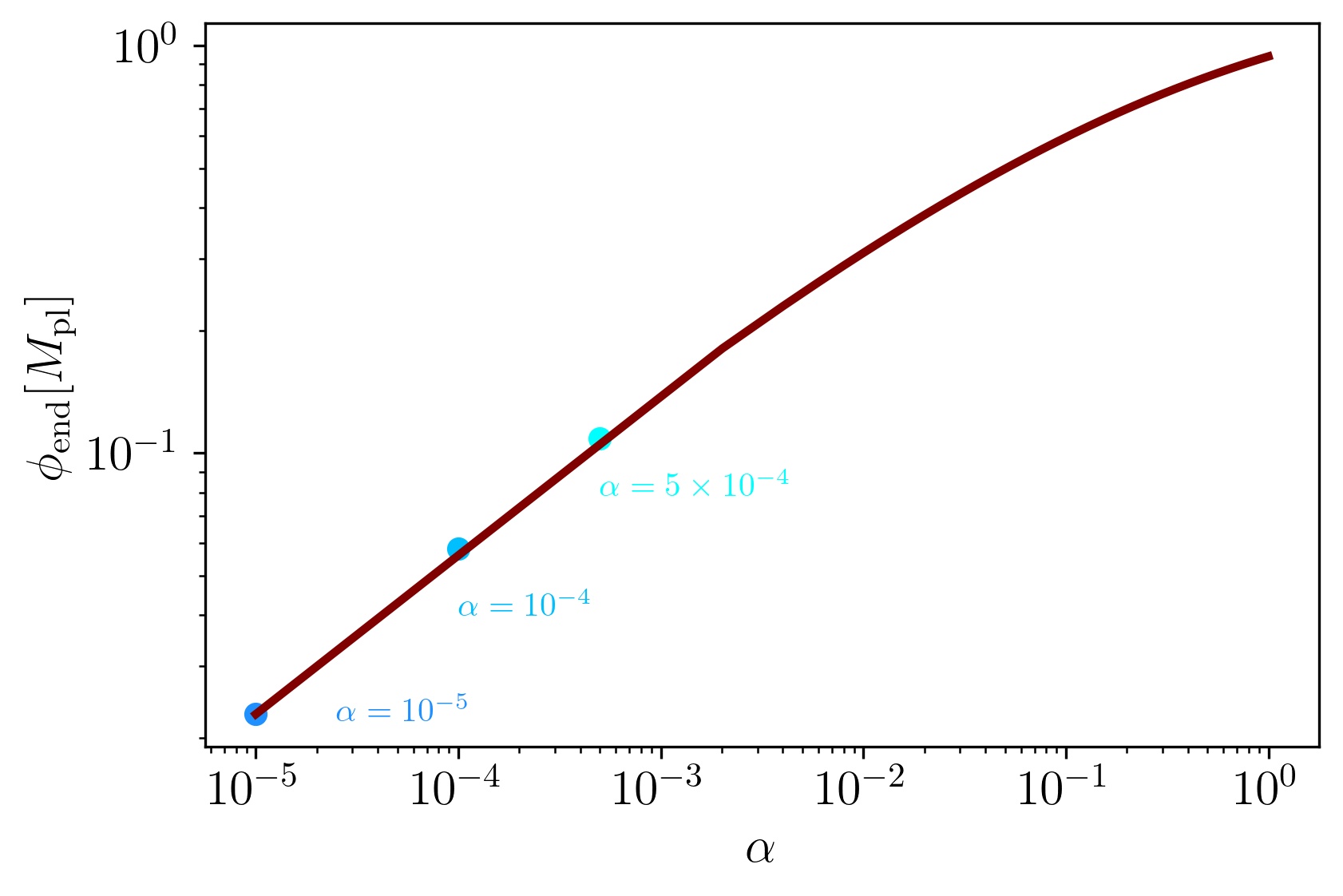}
    \caption{Variation of $\phi_\text{end}$ as a function of $\alpha$. The three points on the plot indicate the different values of $\alpha$ that were used in the lattice simulations.}
    \label{fig:my_label}
\end{figure}
The mass scale $m$ is determined through the CMB normalization at the pivot scale $k_\star$ which requires the field value $\phi_\star$ for some prescribed number of observable $e$-folds. In the slow-roll regime, the total number of $e$-folds is given by
\begin{align}
    N(\phi_\star)&=\frac{1}{M_\text{pl}^2}\int_{\phi_\text{end}}^{\phi_\star}\frac{V}{\partial_\phi V}\dd \phi \\
                 &=\frac{1}{4}\left[ 3\alpha\left( e^{\sqrt{\frac{2}{3\alpha}}x_\star }-e^{\sqrt{\frac{2}{3\alpha}}x_\text{end} } \right)-\sqrt{6\alpha}\left( x_\star-x_\text{end} \right) \right]
\end{align}
where $x=\frac{\phi}{M_\text{pl}}$ and $x_\text{end}$ is determined using Eq. \eqref{eq:phi_end}. Using $55$ $e$-folds as a reference number for observable $e$-folds, the value of $\phi_\star$ can be determined as a function of $\alpha$. Finally, the value of $m$ can be set using the CMB normalization of the primordial power spectrum at $k_\star$. In slow-roll
\begin{equation}
    \mathcal{P}_{\zeta\star}\approx \frac{V^3}{12\pi^2\left( \partial_\phi V \right)^2 M_\text{pl}^6}
\end{equation}
such that
\begin{equation}
    \frac{m}{M_\text{pl}}\approx\sqrt{\frac{128\pi^2\mathcal{P}_{\zeta\star}}{3\alpha^2}}\frac{z_\star}{\left( z_\star - 1 \right)^2}
\end{equation}
where $\mathcal{P}_{\zeta\star}=2.1\times 10^{-9}$ and $z_\star=\exp\left( \sqrt{\frac{2}{3\alpha}}\frac{\phi_\star}{M_\text{pl}} \right)$.

\section{Physical sizes of overdensities}\label{sec:proper_size}
\begin{figure}[h]
    \centering
    \includegraphics[scale=0.45]{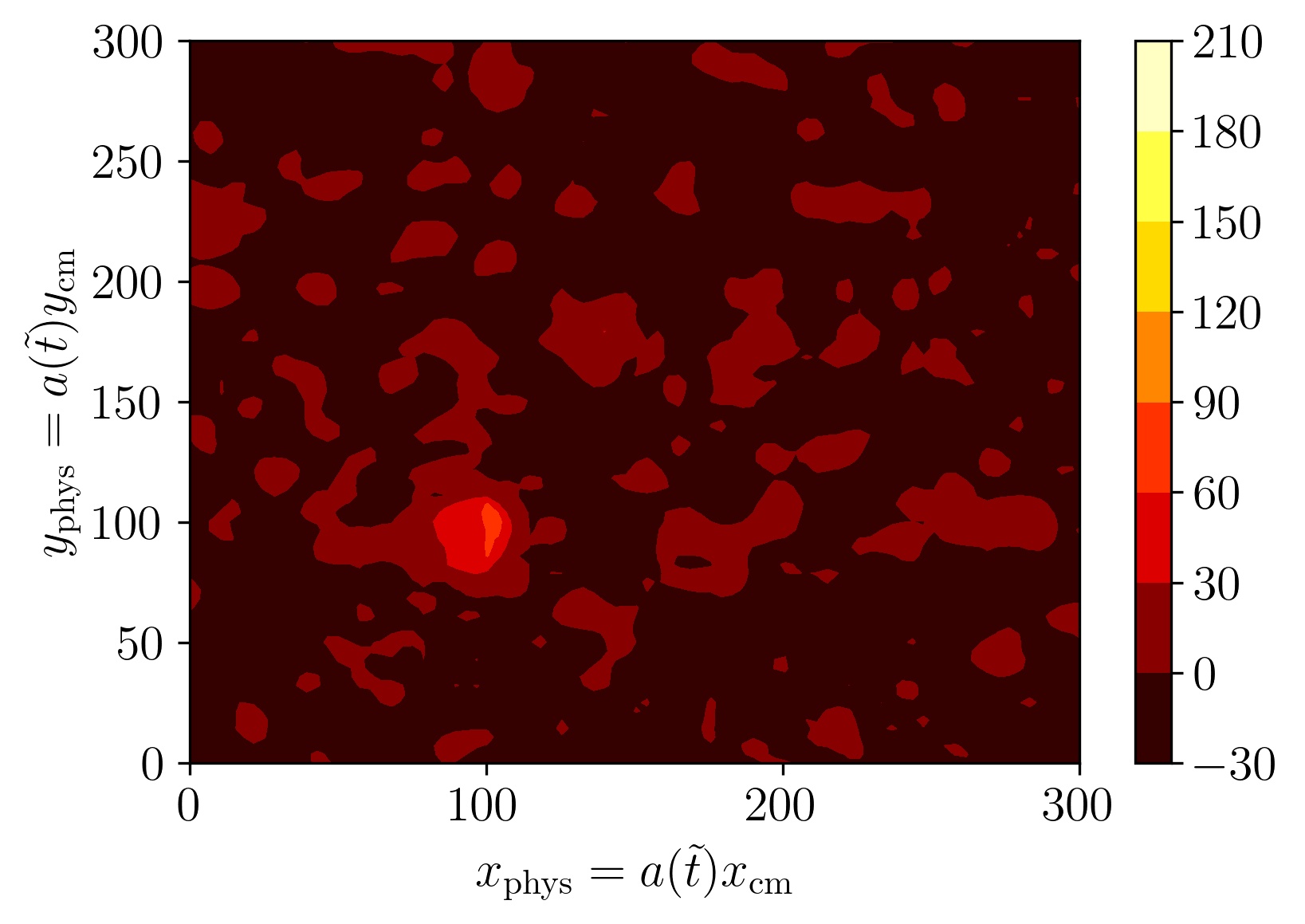}
    \includegraphics[scale=0.45]{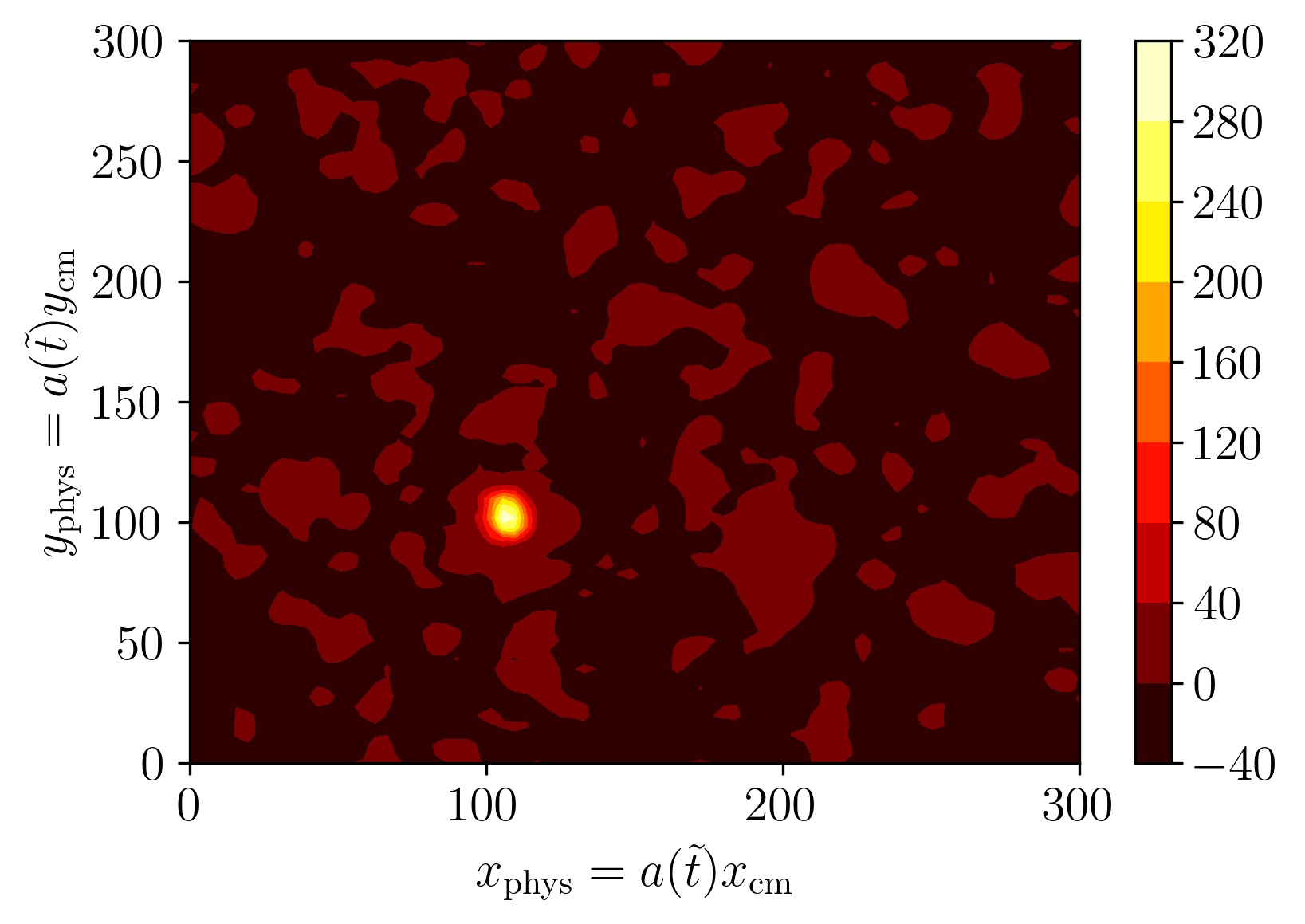}
    \includegraphics[scale=0.45]{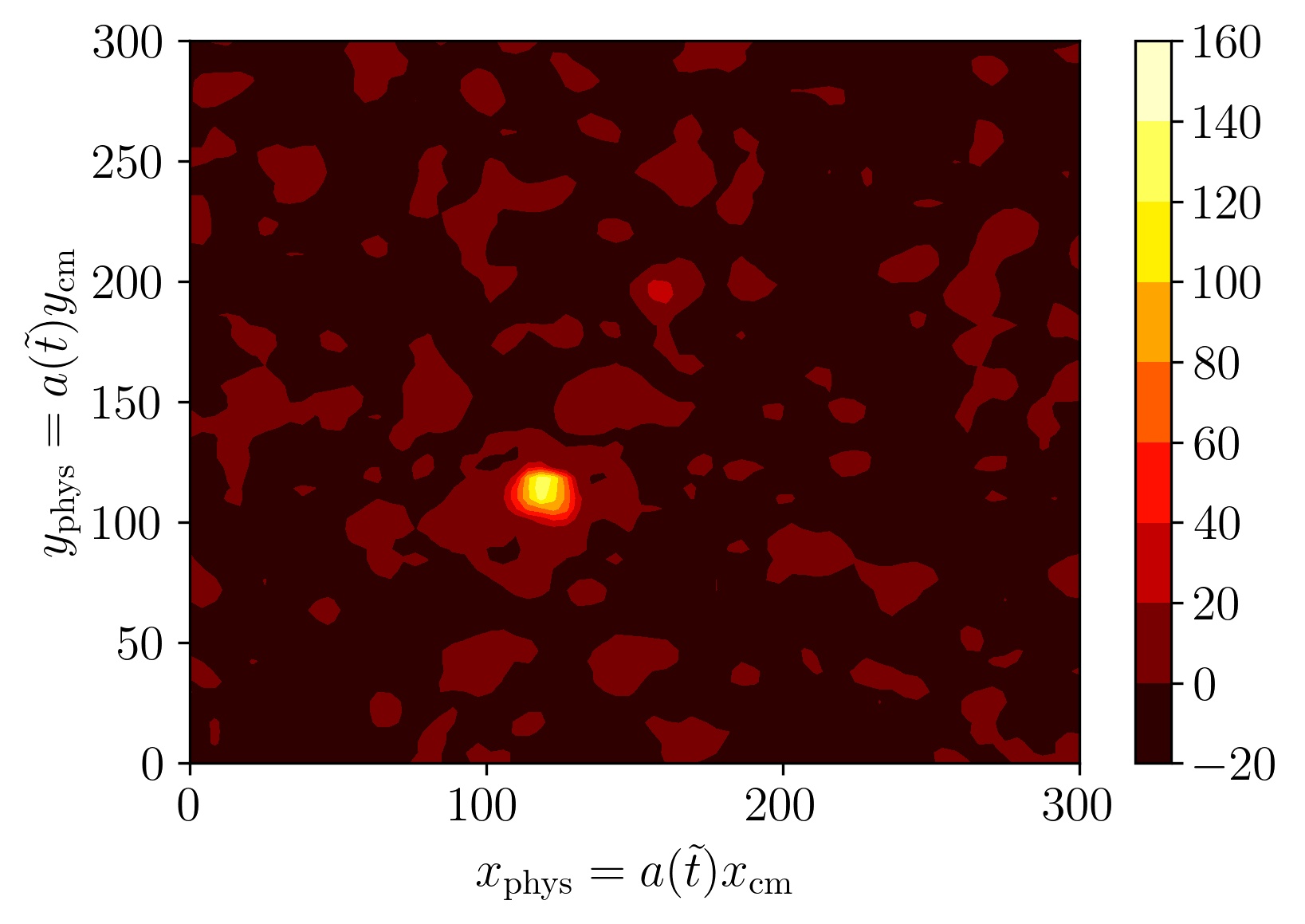}
    \includegraphics[scale=0.45]{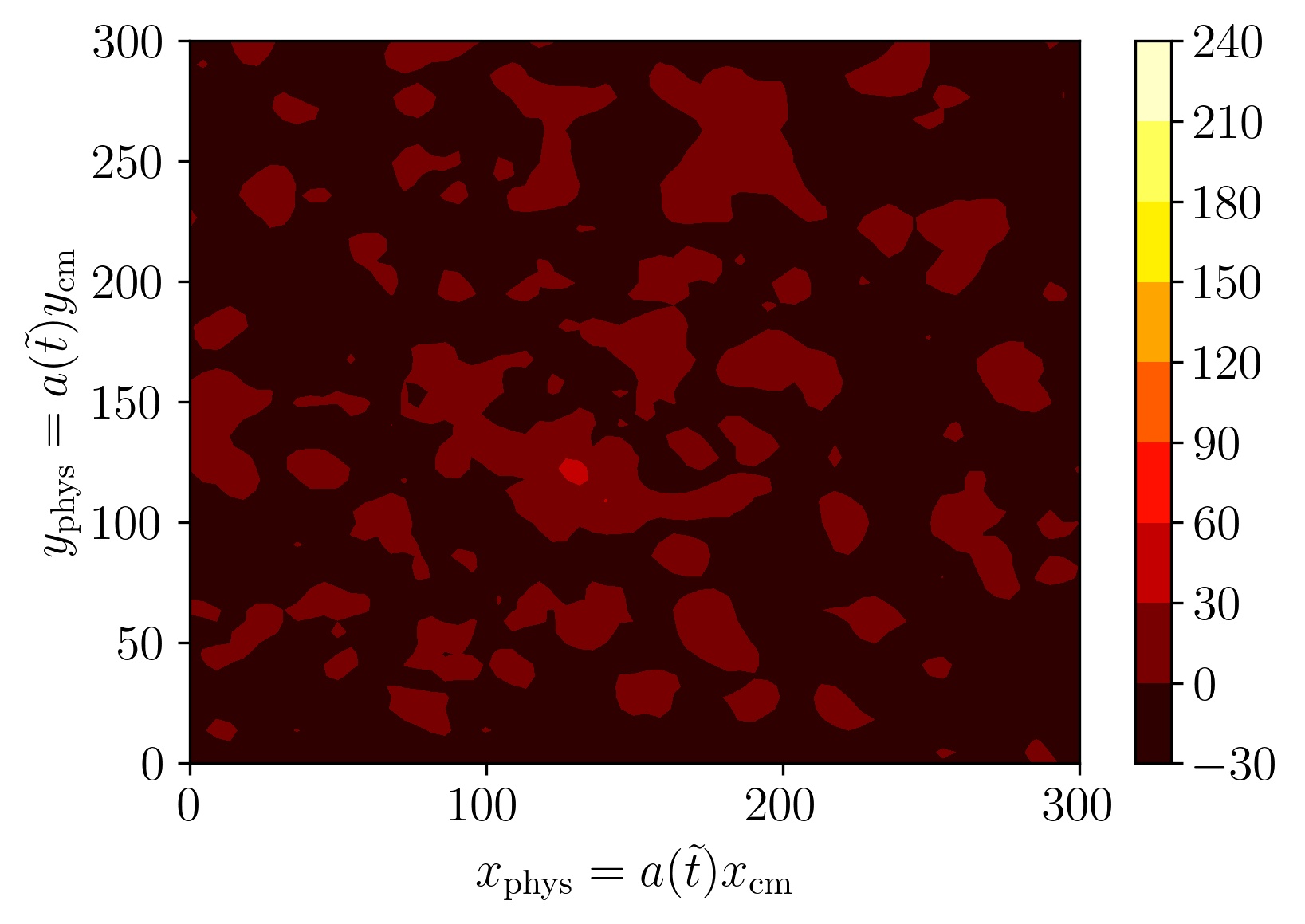}
    \caption{$2d$ slices of the $3d$ energy isosurfaces are shown here for $\alpha=10^{-4}$ at different times, where $\tilde{t}=600$ (top left), $700$ (top right), $800$ (bottom left) and $900$ (bottom right) respectively. }
    \label{fig:slices}
\end{figure}

In Sec. \ref{sec:pspectra_and_structure_formation} we mentioned that the nonlinearities that develop maintain constant physical sizes. This can be demonstrated by using the fact that
\begin{equation}
    \tilde{L}_\text{phys}(\tilde{t})=a(\tilde{t})\tilde{L}=\left( \frac{\tilde{t}}{\tilde{t}_0} \right)^{2/3}\tilde{L}
\end{equation}
where the appropriate scale factor for the matter-dominated epoch has been used. In Fig. (\ref{fig:slices}), the $2d$ slices of the overdensities are shown for $\alpha=10^{-4}$ at four different times where we have zoomed into a $300\times300$ subset of the overall grid. In the plots, the lengths of the grids have been scaled to reflect the physical sizes of the overdensities and it can be seen that, on average, the nonlinear patches maintain roughly constant physical sizes.

\bibliographystyle{apsrev4-1}
\bibliography{refs}

\end{document}